\documentclass[aps,pra,onecolumn,amsmath,notitlepage, showpacs,superscriptaddress, 10pt, longbibliography]{revtex4-1}  

 \pdfoutput=1  

\usepackage{graphicx}  
\usepackage{bm}        
\usepackage{amssymb}   
\usepackage[caption=false]{subfig} 
\usepackage{units} 
\usepackage{epstopdf, array}
\usepackage{hyperref}
\usepackage{color} 
\usepackage{tikz}
\usepackage[percent]{overpic}

\hypersetup{
    pdftitle = {Clark_Symmetry-constrained electron vortex propagation},
    pdfauthor = {Clark et al.}
}

\begin{document}

\renewcommand{\tablename}{FIG.}
 \renewcommand{\thefigure}{\Roman{figure}}

\title{Symmetry-constrained electron vortex propagation}

\author{L. Clark}
\email[Electronic address: ]{laura.clark@uantwerp.be}
\affiliation{EMAT, University of Antwerp, Groenenborgerlaan 171, 2020 Antwerp, Belgium}
\author{G. Guzzinati}
\affiliation{EMAT, University of Antwerp, Groenenborgerlaan 171, 2020 Antwerp, Belgium}
\author{A. B\'ech\'e}
\affiliation{EMAT, University of Antwerp, Groenenborgerlaan 171, 2020 Antwerp, Belgium}
\author{A. Lubk}
\affiliation{Triebenberglabor, University of Dresden, Zum Triebenberg 1, 01062 Dresden, Germany}
\author{J. Verbeeck}
\affiliation{EMAT, University of Antwerp, Groenenborgerlaan 171, 2020 Antwerp, Belgium}

\date{\today}

\begin{abstract}
Electron vortex beams hold great promise for development in transmission electron microscopy, but have yet to be widely adopted.
This is partly due to the complex set of interactions that occur between a beam carrying orbital angular momentum (OAM) and a sample.
Herein, the system is simplified to focus on the interaction between geometrical symmetries, OAM  and topology. We present multiple simulations, alongside experimental data to study the behaviour of a variety of electron vortex beams after interacting with apertures of different symmetries, and investigate the effect on their OAM and vortex structure, both in the far-field and under free-space propagation.
\end{abstract}

\maketitle

\section{Introduction}
\setcounter{table}{1}

Electron vortex beams, are electron states wherein the phase is structured to possess a phase singularity on axis, with a wave function of $\Psi=A(r) \mathrm{e}^{i \ell \phi} \mathrm{e}^{i k z}$. The amplitude function in this simple case,  is a function of $r$ alone. The paraxial beam propagates in the $z$-direction, with wave-number $k$. The value $\ell$ is an integer, describing the topological charge, or winding number of the vortex. When the beam is in a simple,  circularly symmetric state, $\ell$ also determines the orbital angular momentum (OAM) of the beam \cite{allen1992orbital}.

Such electron beams were first considered explicitly in 2007 \cite{bliokh2007semiclassical}, having been alluded to previously \cite{bialynicki2000motion, fukuhara1983electron}. They were demonstrated experimentally in 2010 \cite{uchida2010generation, verbeeck2010production, mcmorran2011electron}, and have remained under intense study since. The interest in these structured beams comes not only from a fundamental interest in their unusual behaviours \cite{van2013spin, van2014rutherford, van2015inelastic, lloyd2012interaction, lloyd2012electromagnetic, bliokh2011relativistic,  mendis2015dynamic, hasegawa2013young}, but also from prospects for novel applications in a variety of topics. Potential applications include magnetic chiral dichroism measurements \cite{verbeeck2010production, schattschneider2014magnetic, verbeeck2011atomic, rusz2013boundaries}, being used to determine crystal chirality \cite{juchtmans2015using} and providing opportunities to use electron vortices as an analogue of optical tweezers \cite{verbeeck2013manipulate}.

The application most fervently awaited for electron vortex beams is that they may enable atomic-resolution electron magnetic circular dichroism (EMCD) \cite{verbeeck2010production, rusz2013boundaries, schattschneider2014magnetic}, with better resolution than existing methods.
While early experiments seemed encouraging, later studies revealed the complexity of the set-up, thus requiring further careful attention
 \cite{rusz2014scattering, rusz2014achieving, pohl2015electron}.

An electron beam interacts strongly with a sample \cite{henderson1995potential}, which rapidly changes the shape of the wavefront.
For typical TEM sample thicknesses, multiple scattering is likely to occur, further complicating the interaction between the beam electron and the sample.
If the beam is aligned with a crystal zone-axis of the sample, the interaction is particularly strong, and channelling will occur, biasing which parts of the sample are exposed to the probing electron \cite{lugg2015quantitativeness}.
This channelling will also affect the propagation of the vortex core \cite{lubk2013topological, loffler2012elastic}.
The local symmetry imposed on the beam by the crystal structure will also affect the OAM of the wave as it propagates \cite{molina2001management}.
All of these effects occur synchronously, leading to complicated details in the electron wave as it exits from the sample, and complex changes in any recorded electron energy-loss spectrograms (as these depend on the local electron wave-function).

Here, we aim to clarify one aspect of this complex system, by studying the effects of spatial confinement in a geometric symmetry on an electron vortex beam.
This is loosely analogous to the effects of the crystal symmetry on the electron vortex beam, while avoiding the complexity of channelling, and other sample interaction effects.

Futhermore, in previous work we noted interesting behaviour of electron vortex beams, when microscope apertures are misaligned, or when using non-round apertures \cite{lubk2013topological, guzzinati2014measuring, clark2014quantitative}. 
Particularly, following the interaction with apertures with discrete rotational-symmetries as in \citeauthor{guzzinati2014measuring} \cite{guzzinati2014measuring}, it can be seen that the original beam symmetry is not conserved, and a dark vortex core is not always found on axis.
The interference between the aperture edge waves \cite{keller1962geometrical} and the vortex probe, leads to an intensity profile which reveals the topological charge of the input beam, in the case of a centred, single-ringed input vortex probe \cite{hickmann2010unveiling}.

Such a non-circularly symmetric constraint prevents the beam from being in an OAM eigenstate, so a high-order vortex core may degenerate into an arrangement of multiple, displaced $|\ell|=1$ vortices \cite{lubk2013topological}.
We now deliberately apply these unusual settings to allow a deeper study of vortex splitting, propagation and  reconnection.

We will investigate the vortex structure of an electron vortex beam after impinging on an aperture of a chosen geometry (round, square, equilateral triangle, off-centred circle), in the far-field, and during propagation.
We will study this system both computationally and experimentally.
Similar systems have been considered previously with optical vortex beams, but only with regard to the far-field behaviour of the beam. Here, we focus on the beam structure during propagation, as a route to a deeper understanding of electron vortices within a sample.
In the following, we firstly compare the effects of the different geometric apertures on the far-field wave and vortex structures, comparing experimental data to our simulations.
We then model the whole focal series of the beam, tracking the propagation of the vortex cores through the volume. We find a surprisingly complex vortex structure, which may be of significance to the development of the vortex-EMCD method.

\section{System Description}

Many different methods of producing electron vortex beams have been investigated. The first experimental electron vortex beam relied on an opportune stacking of graphite flakes \cite{uchida2010generation}. This is not easily reproducible. 
After that, the holographic forked mask technique, borrowed from optics \cite{bazhenov1990laser},  was introduced \cite{verbeeck2010production}, and developed  further \cite{mcmorran2011electron, verbeeck2012new, clark2012electron, grillo2014highly, harvey2014efficient}. 
Electron vortices produced through aberration manipulation were also investigated \cite{petersen2013electron, clark2013exploiting, schattschneider2012novel, tavabi2015tunable}, as well as direct phase plate production \cite{shiloh2014sculpturing, beche2016focused}.  These methods however, all have their own foibles - primarily, limitations to the intensity in a given vortex order, or experimental limitations to the attainable purity of a given OAM state. 

More recent developments however, have improved on this. It has been shown that illuminating the end of a long, thin, nanoscale bar magnet, can induce a vortex character into the electron beam, while only blocking a very small proportion of the incoming electron flux \cite{beche2014magnetic, blackburn2014vortex}. Experimentally, if the magnetic needle is produced sufficiently accurately, the resultant vortex beam can have more than $95\%$ of the beam's intensity in the desired OAM state \cite{beche2014communication}.
While this method results in the ``best'' electron vortex beams currently available, it is not well suited to investigating different orders of topological charge, as this would require access to a set of several magnetic needle apertures.
To ensure the fairest comparison between a set of vortex beams of different order, we will thus investigate the beams resulting from a well-made forked holographic mask, ensuring they are sufficiently laterally separated.
This has the benefit of having high OAM-mode purity, at the expense of lower beam intensity, not critical in the present study.

\begin{figure*}
\subfloat[Simple forked vortex mask design, with bar width identical to gap width.]{\label{fig:basicForkedMaskDiagrama}\includegraphics[width=0.3\linewidth]{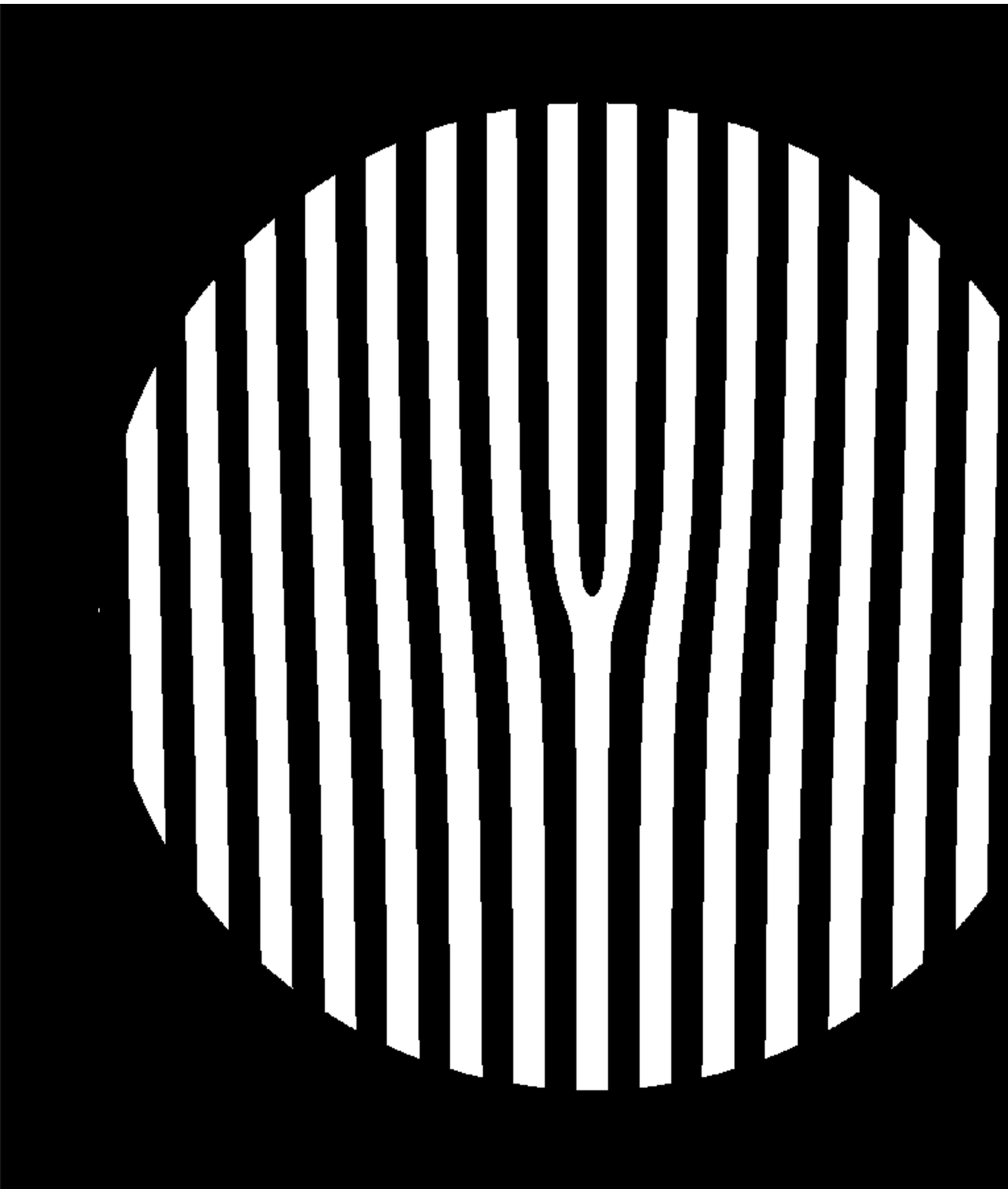}} 
\qquad \qquad \qquad \qquad
\subfloat[Experimental forked mask, $10 \mu$m diameter. Note that the gaps are slightly wider than the bars.]{\label{fig:basicForkedMaskDiagramb}\includegraphics[width=0.3\linewidth]{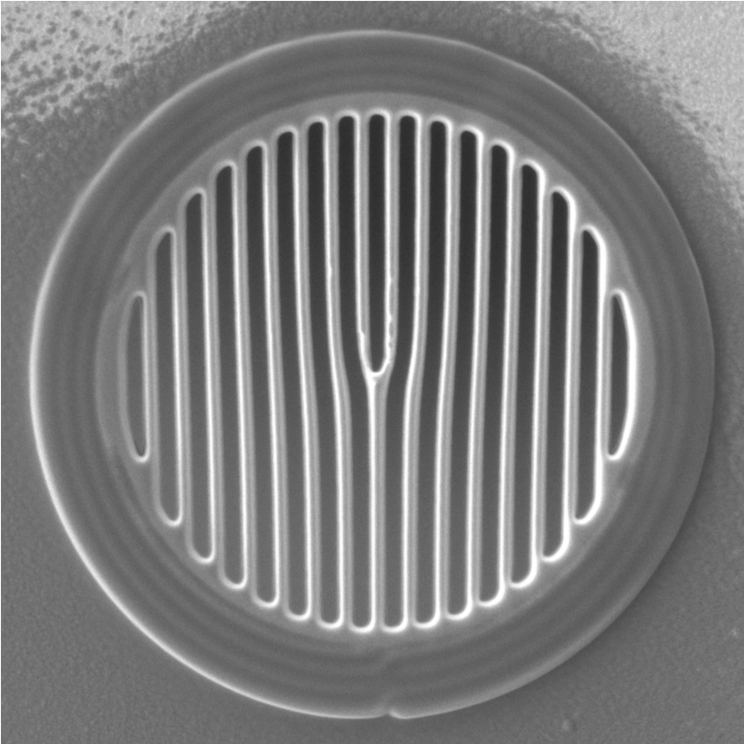}} 
\caption{Theoretical and experimental comparison of forked mask apertures.}
 \end{figure*}

By using a forked holographic mask in the condenser plane of the microscope, we are able to produce a range of high-order vortex beams in the image plane, with diameters in the range of a few tens of nanometres (adjustable by tuning the condenser system).
In the ideal, simple case, where the mask is designed by interfering a tilted plane wave with an $\exp (i \ell \phi)$ term only (with no additional amplitude modulation), and the binarisation is applied at exactly half of the maximum intensity value, the bars in the resulting mask will be the same width as the gaps (as illustrated in figure \ref{fig:basicForkedMaskDiagrama}). In this case, the only beams produced will be the zero-order central beam (with $\ell =0$), and odd-ordered side beams (with $| \ell |=1, 3, 5...$), due to the combination of diffraction envelopes creating missing orders \cite{hecht1998hecht}.

However, in many experimental cases, this does not hold true, due to the challenges in the fine control of focused ion beam (FIB) milling time \cite{clark2012electron}, leading to other vortex orders appearing, at varying intensities.
The milling of our present experimental mask enables access to $|\ell |= (0,1,2,3)$ vortex orders with reasonable intensities.

For maximum diffractive effects, we want to produce the geometric apertures such that they are of a similar size to the vortex beams at the sample plane.
Producing apertures at this scale is challenging; we make use of the conjugated selected area plane instead.
There is a demagnification factor of approximately $70$ between the selected area plane in our microscope, and the image plane, allowing us to use apertures with a characteristic dimension of $2 \mu$m. This is projected to the image plane as a characteristic dimension of $30$~nm, a similar size to the vortex beams we produce from a forked mask with diameter $10$~$\mu$m (approximately $0.25$~mrad) when slightly defocused to optimally fill the apertures.

In this way, we can illuminate each of the geometric apertures with electron-vortex beams of different topological charges. We can image the far-field diffraction of each, and compare with simulations to investigate the resulting vortex structure. 
Furthermore, we can also manipulate the diffraction and projector lenses, such as to image a full series of the electron-vortex propagation volume, allowing us to study the free-space propagation of a pure vortex state after passing through a specific aperture.
This will show how a pure state splits into a complicated combination of vortex states, depending on the symmetry of the system.

\subsection{Distinguishing topological charge, and orbital angular momentum}
OAM and topological charge are often used interchangeably. They do coincide in the well-studied case of a single, on-axis vortex beam.
However, in the general case, they are different \cite{berry2009optical}.
They can differ in magnitude even in some on-axis, circularly symmetric cases \cite{amaral2014characterization}, or differ in both sign and magnitude \cite{molina2012vortex}.
To aid clarity in the following sections, we explicitly define these concepts here.

Topological charge is identical to winding number and is a local property. It describes the number of times the phase passes through $2 \pi$ around a closed loop, $C$ :
\begin{equation}
\ell= \frac{1}{2 \pi} \oint_C \! \mathrm{d}\phi
\end{equation}
If the loop itself passes through no zero intensity points, $\ell$ is constrained to being an integer. The total topological charge in a region is only affected by the discrete points of phase discontinuity within that region. For an in depth discussion on the nature of topological charge, see the work of Dennis \cite{dennis2001topological}.

OAM is axis-dependent, varying strongly with where in the beam it is measured from. It is only conserved in a circularly symmetric system \cite{ferrando2005discrete}.
Typically, in our paraxial systems, we are only interested in the $z$-component of the expectation value of the OAM:
\begin{equation}
\langle L_z \rangle \, =\,  \langle \Psi | \hat{L}_z | \Psi \rangle
\end{equation}
where, in cylindrical coordinates, $ \hat{L_z}= -i \hbar \frac{\partial}{\partial \hat{\theta}}$ \cite{schattschneider2011theory, molina2001management}. 
In a system with discrete, rotational symmetry, a pure input state will broaden into a superposition of OAM modes upon propagation.
The input vortex core may degenerate into a beam with multiple first-order vortices, displaced from the centre of the incoming beam.
Pairs of $+1, -1$  vortices can also be created, or annihilated within a freely propagating system, forming $3$D vortex loops which do not change the global topology of the system  \cite{lubk2013topological, allen2001phase}.

\subsection{Endurance of geometrical symmetry}
In addition to the propagation behaviour enforced by the conservation of topology, the standard behaviour of Fourier optics must still apply.
It is trivial to show that the intensity pattern in the far-field of a plane wave limited by an aperture, is the squared modulus of the Fourier transform of the wave in the aperture plane \cite{goodman2005introduction}.
Distorting the input wave from a planar wave-front to something else will lead to a convolution of the Fourier transform of the aperture shape, with the Fourier transform of the  input wave. Combining these effects means that the complete rotational symmetry of the incoming vortex beam  is reduced to the discrete rotational symmetry of the aperture.

\section{Far-field behaviour}
We begin investigating these effects, by studying far-field intensities of different vortex beam orders on a set of apertures, with different rotational symmetries.

\subsection{Simulated Results}
\begin{table}
\centering
\begin{tabular}{>{\centering\arraybackslash}m{1.8cm} >{\centering\arraybackslash}m{3cm} >{\centering\arraybackslash}m{3cm} >{\centering\arraybackslash}m{3cm} >{\centering\arraybackslash}m{3cm}}
Apertures  &  \includegraphics[width=3cm]{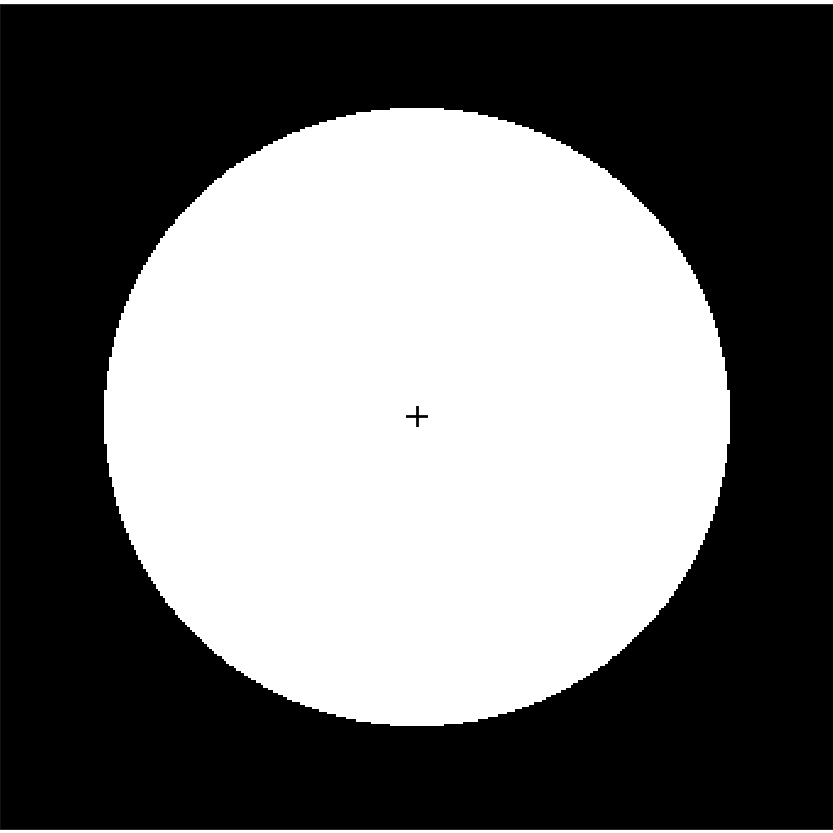}   &   \includegraphics[width=3cm]{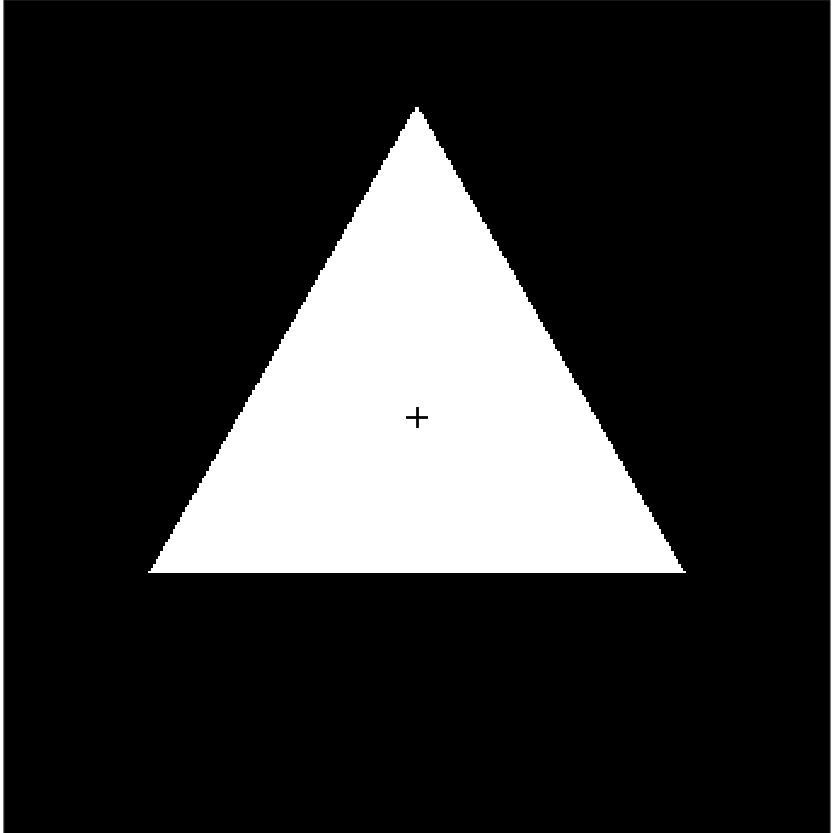}  &   \includegraphics[width=3cm]{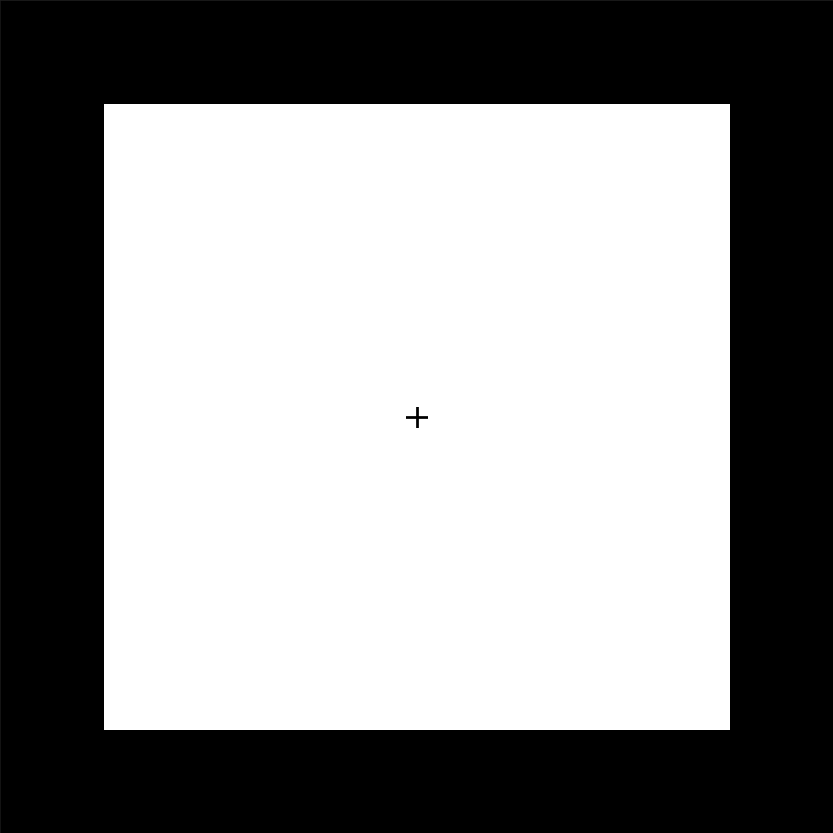} &   \includegraphics[width=3cm]{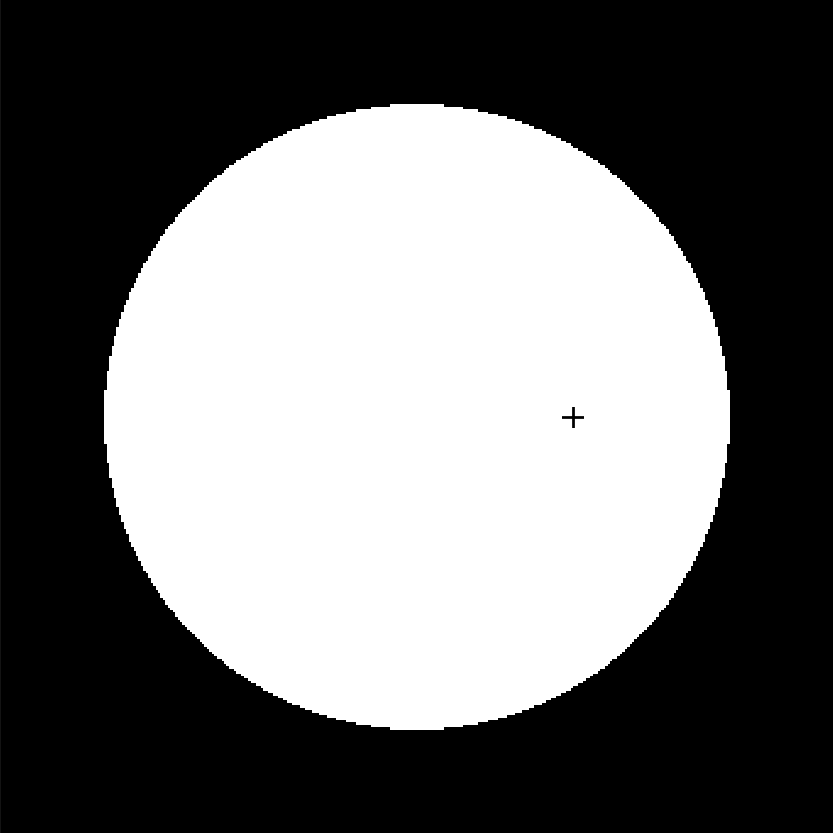}  \\

Input  $ \ell=0$
 \includegraphics[width=1.5cm]{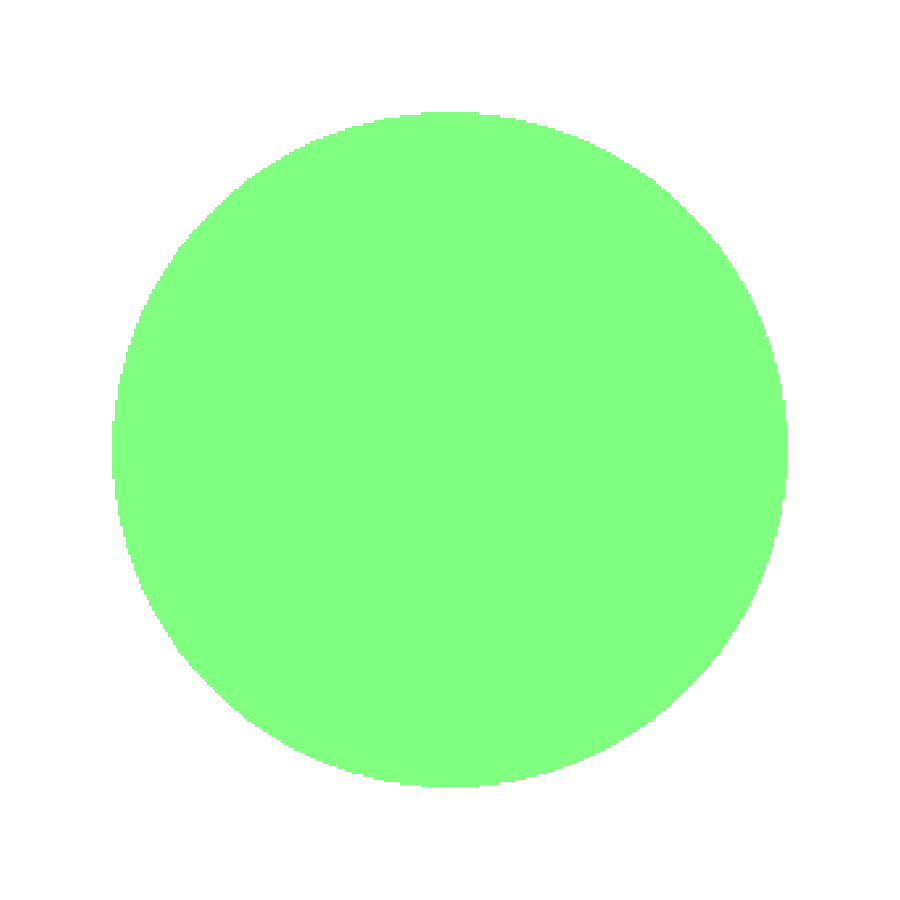} &  \includegraphics[width=3cm]{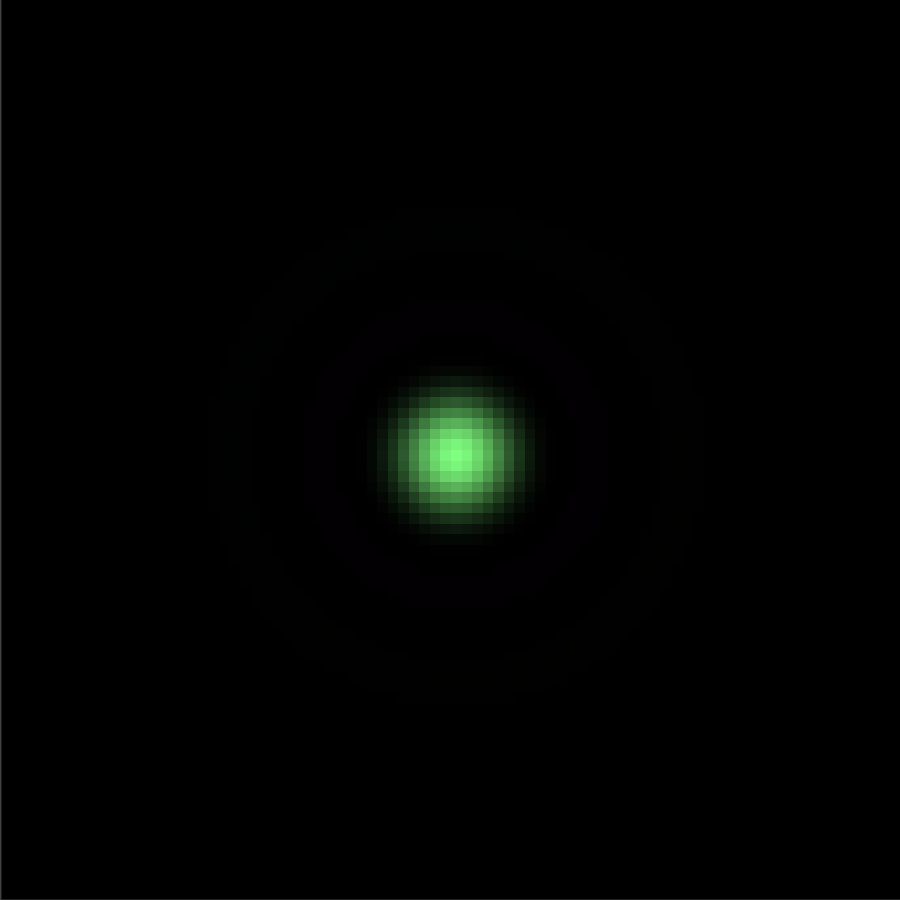}   &   \includegraphics[width=3cm]{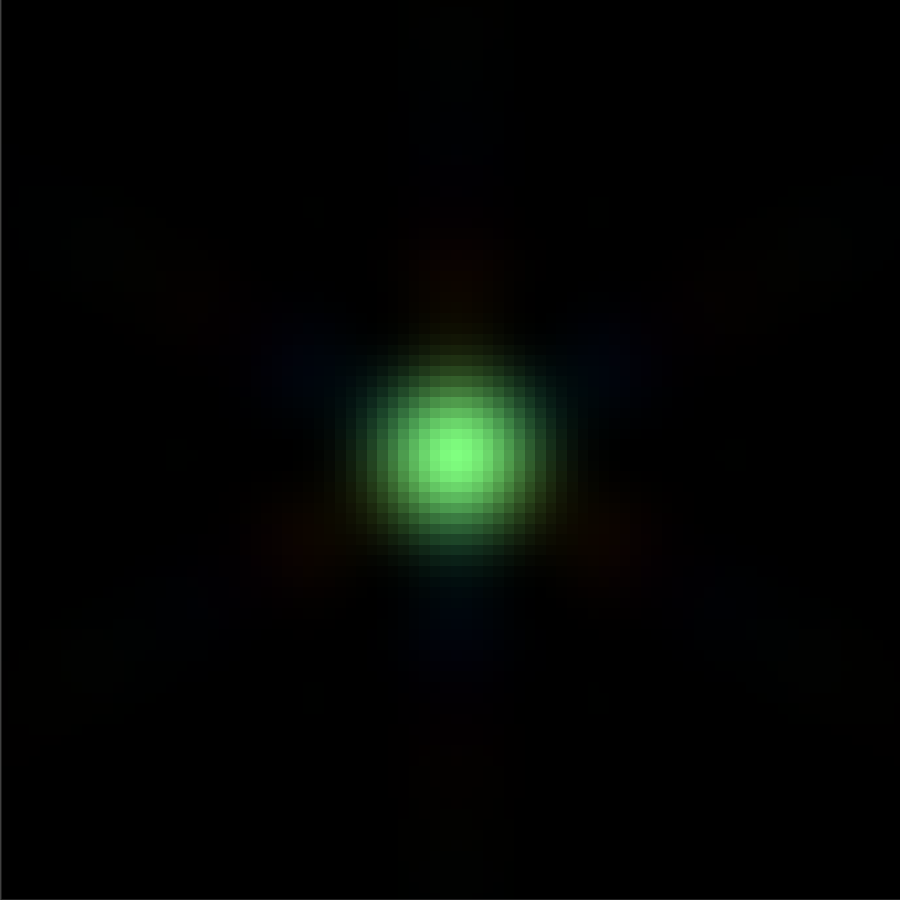}  &   \includegraphics[width=3cm]{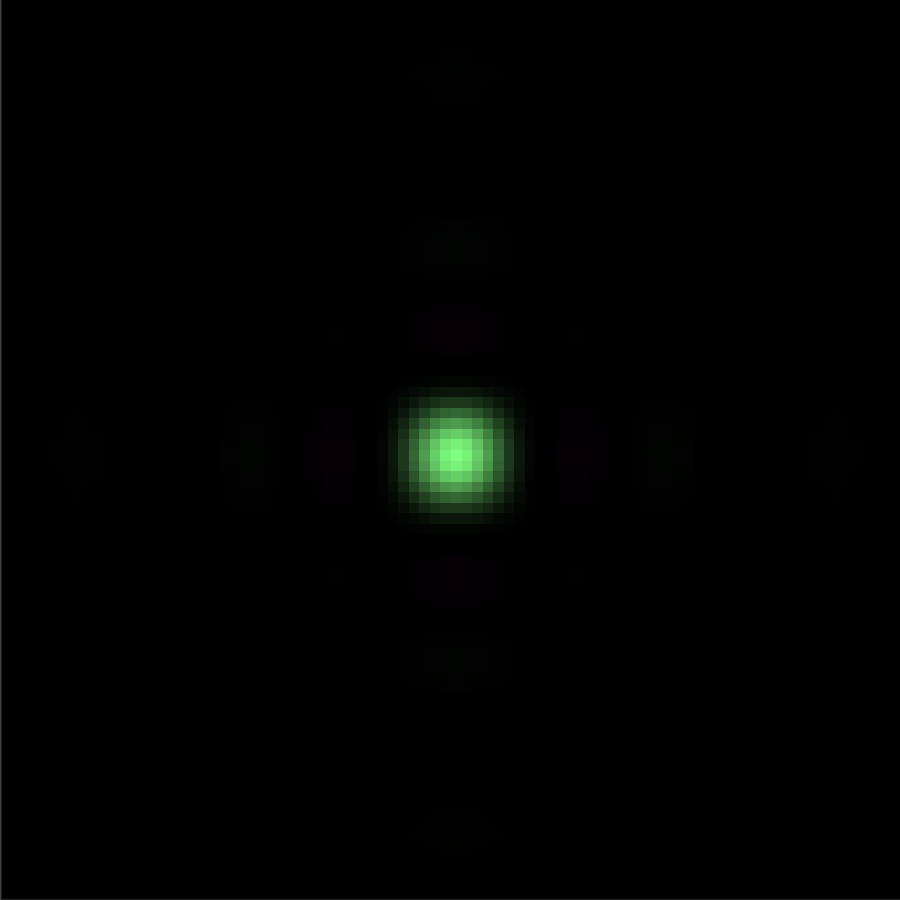}  &  \includegraphics[width=3cm]{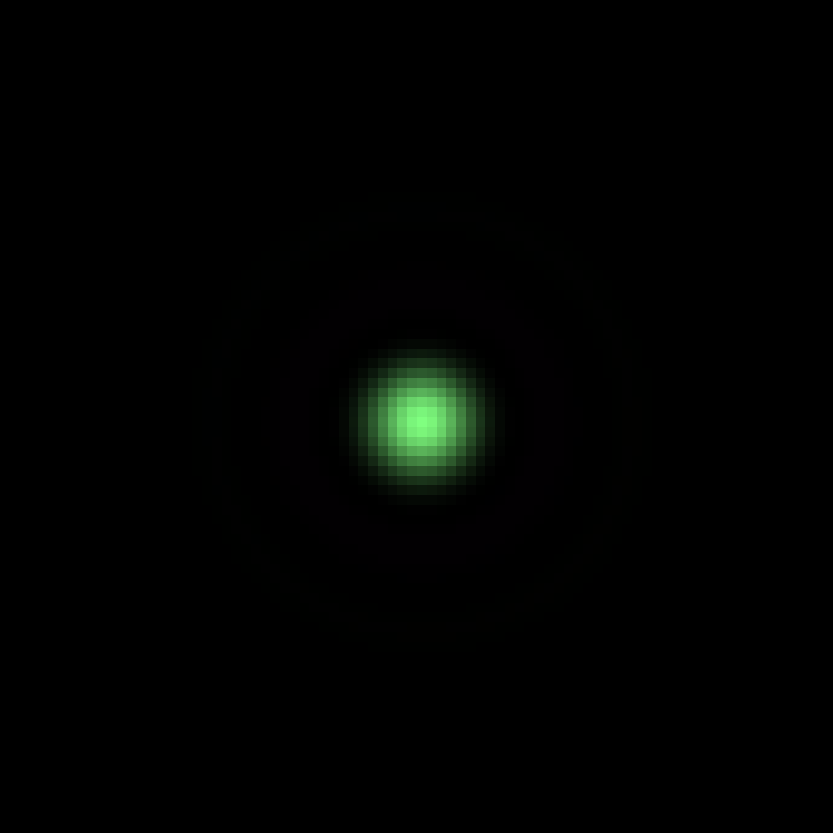}  \\

Input $\ell=1$ 
\includegraphics[width=1.5cm]{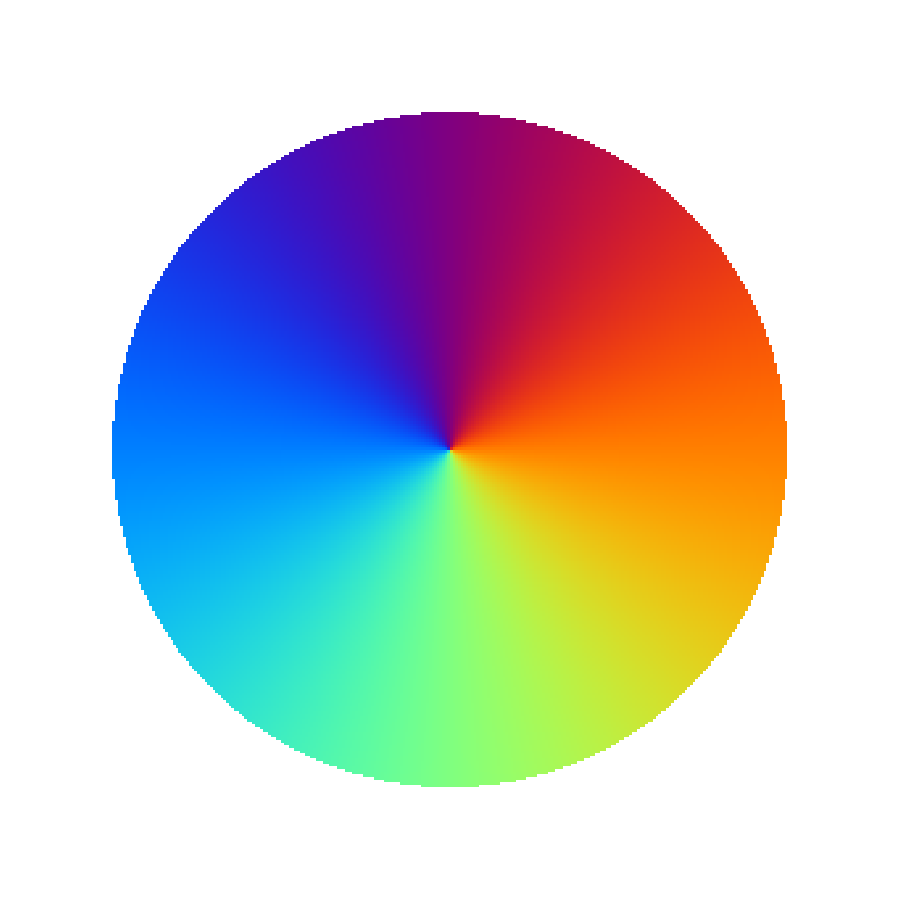}&  \includegraphics[width=3cm]{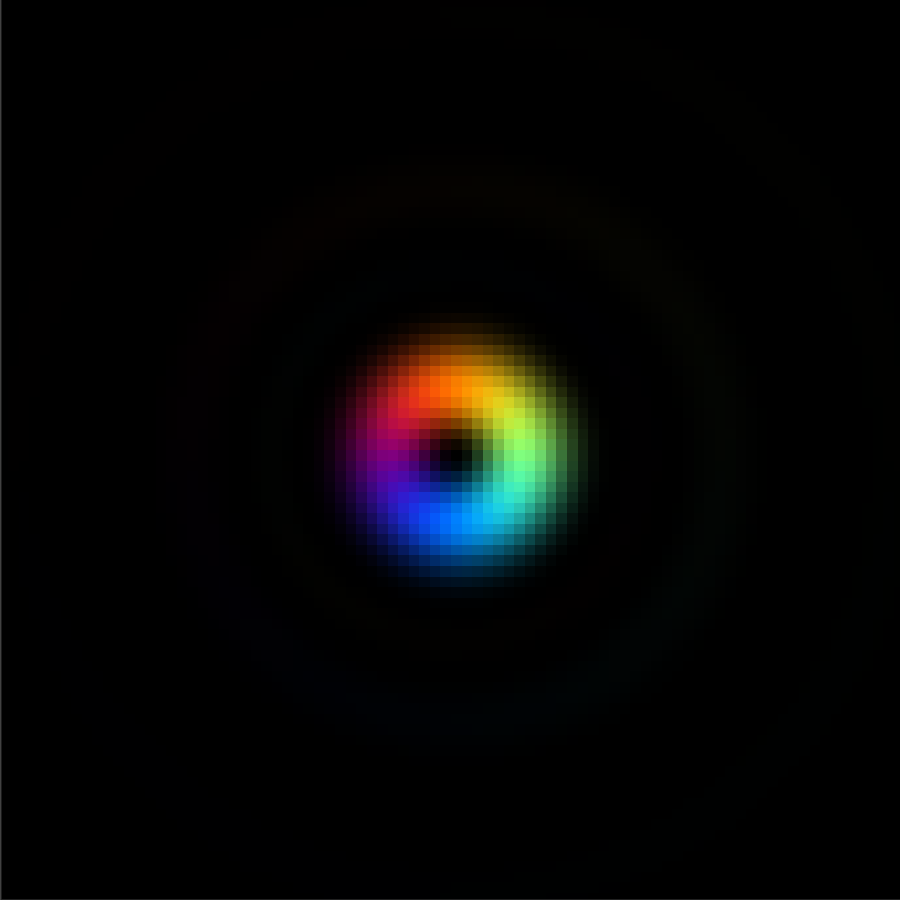}   &   \includegraphics[width=3cm]{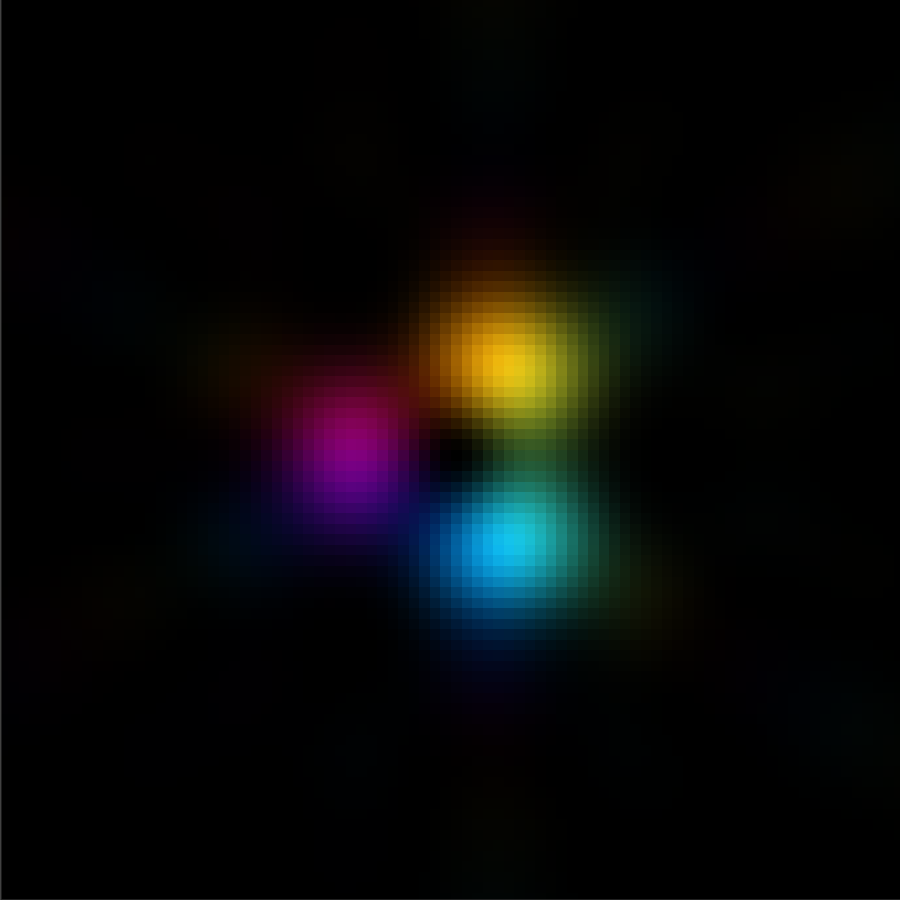}  &   \includegraphics[width=3cm]{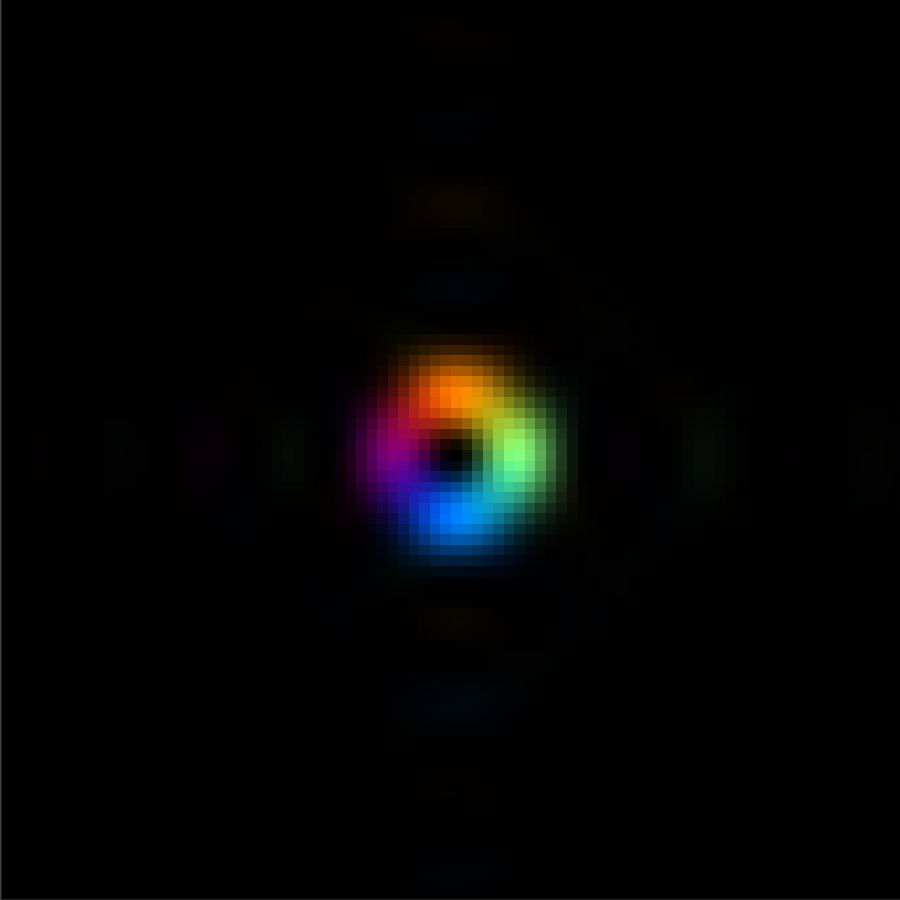}  &  \includegraphics[width=3cm]{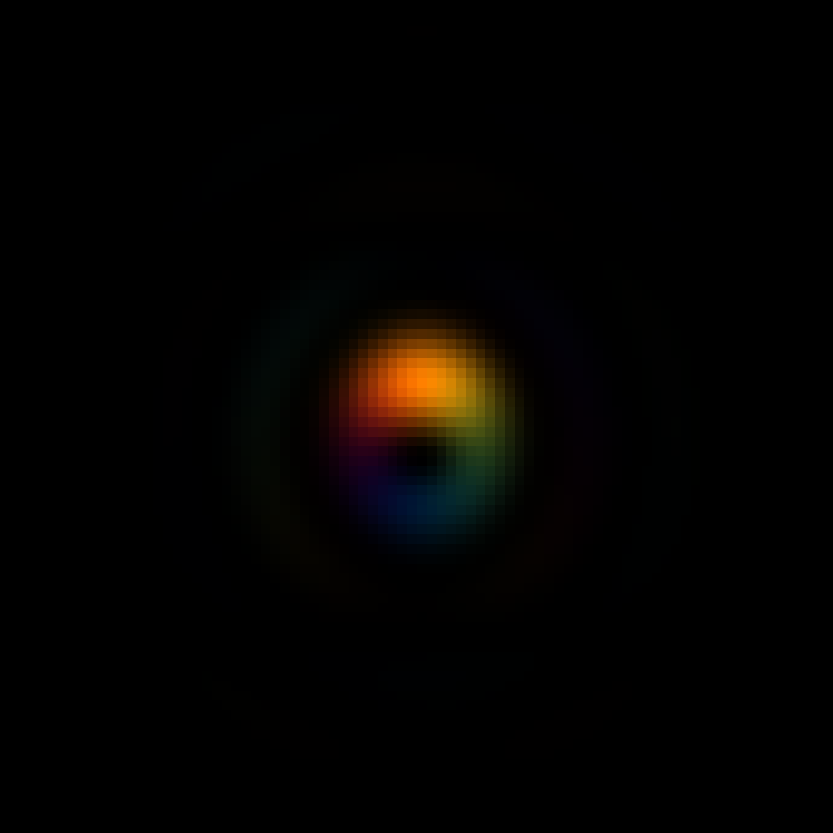}  \\

Input $\ell=2$ 
\includegraphics[width=1.5cm]{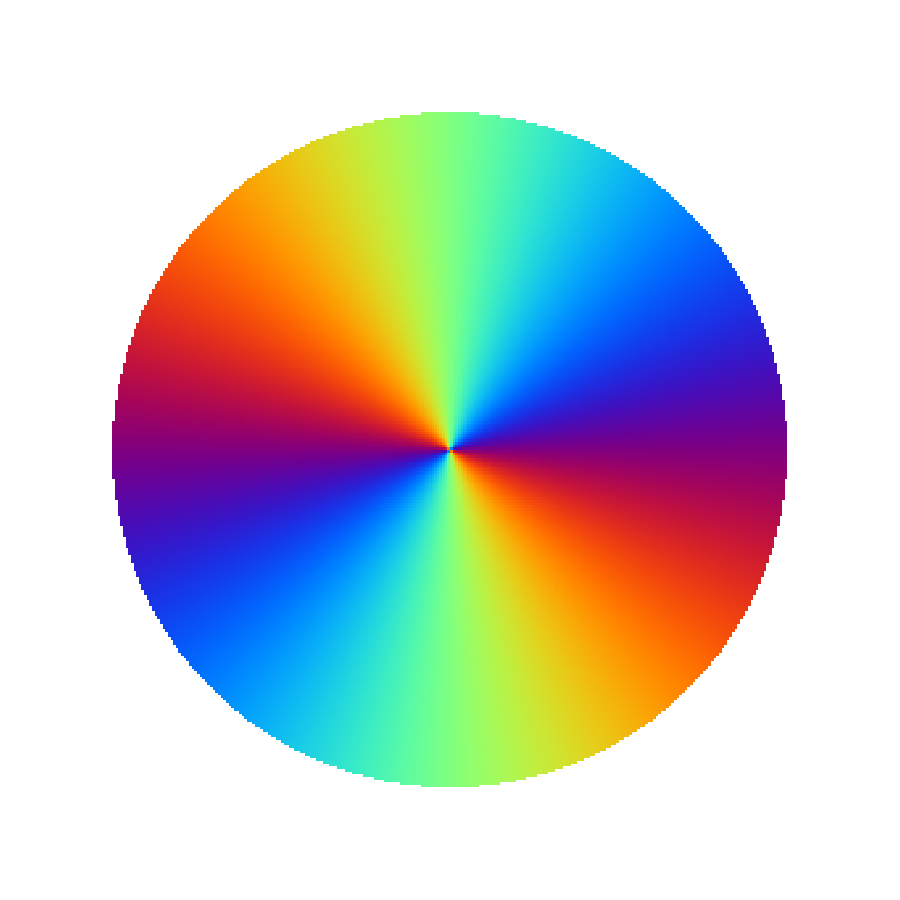}&  \includegraphics[width=3cm]{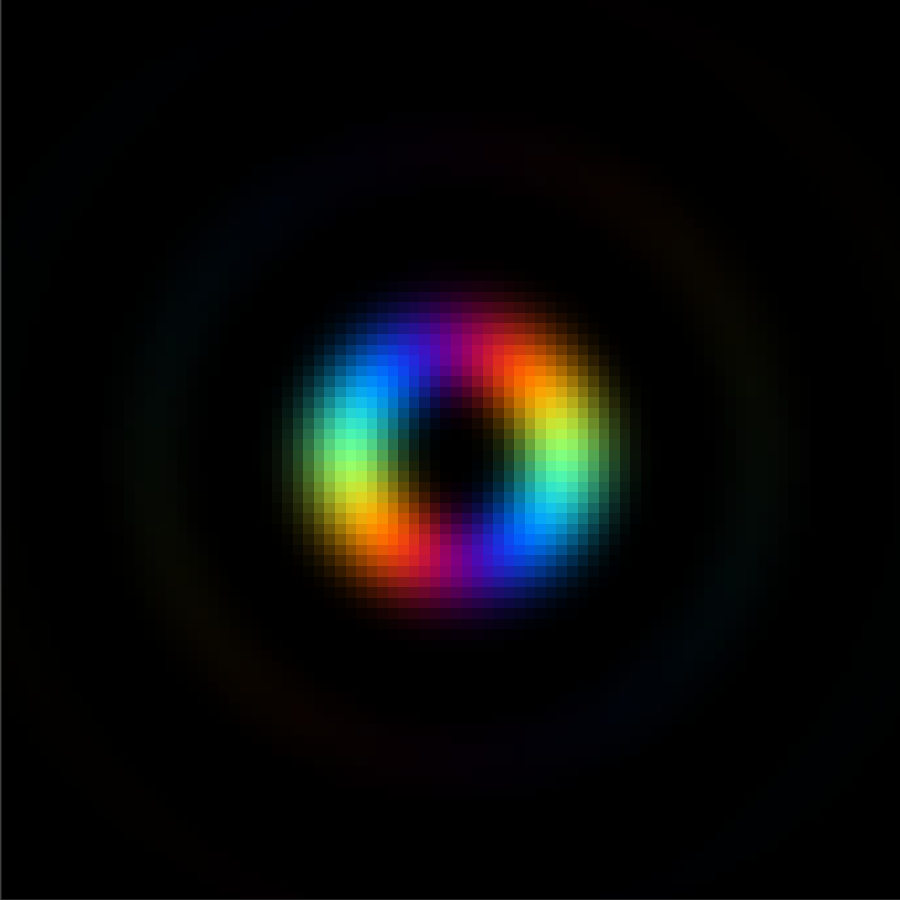}   &   \includegraphics[width=3cm]{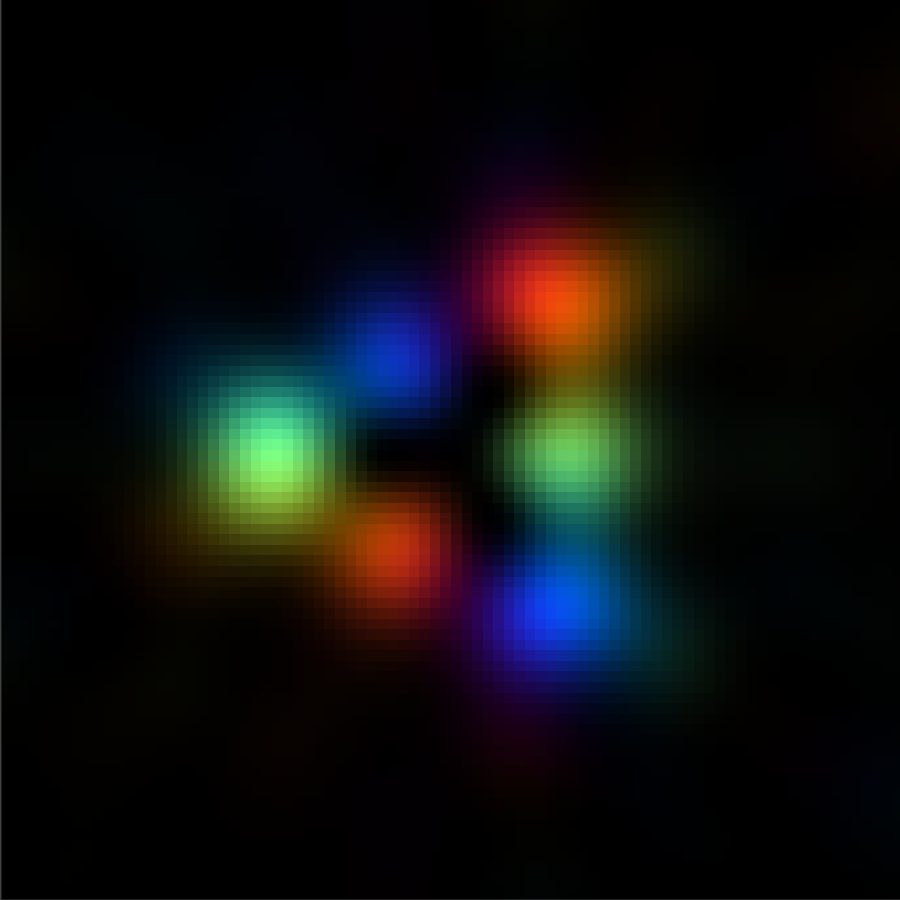}  &   \includegraphics[width=3cm]{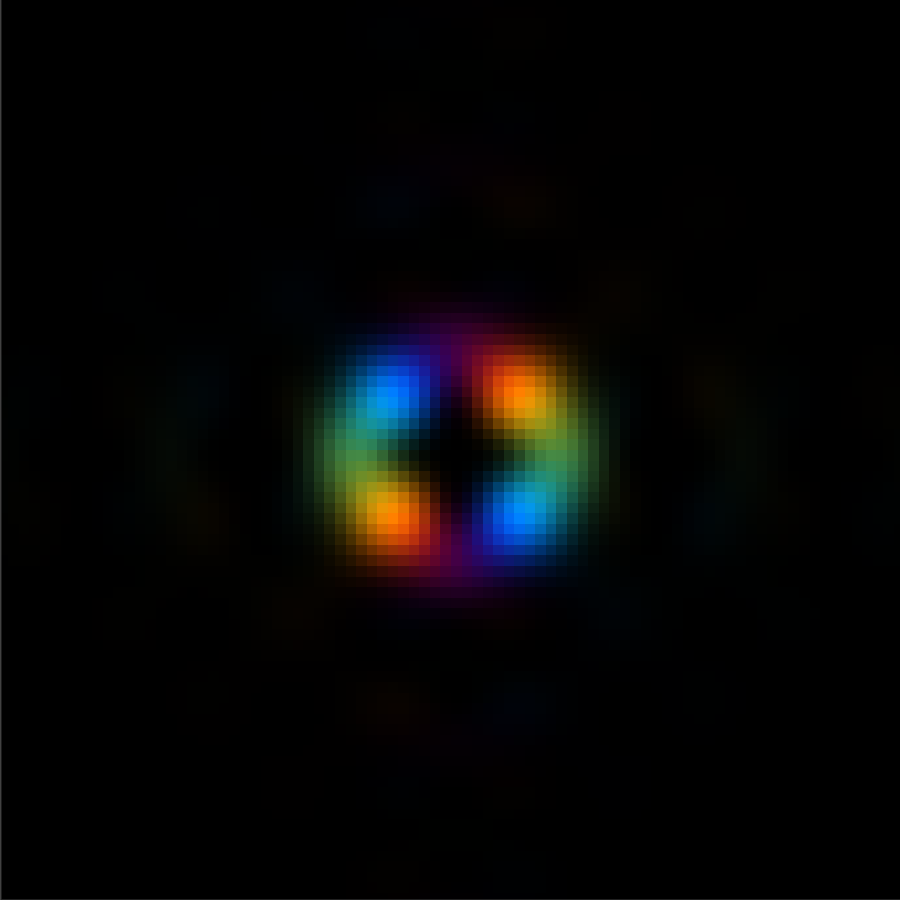}  &  \includegraphics[width=3cm]{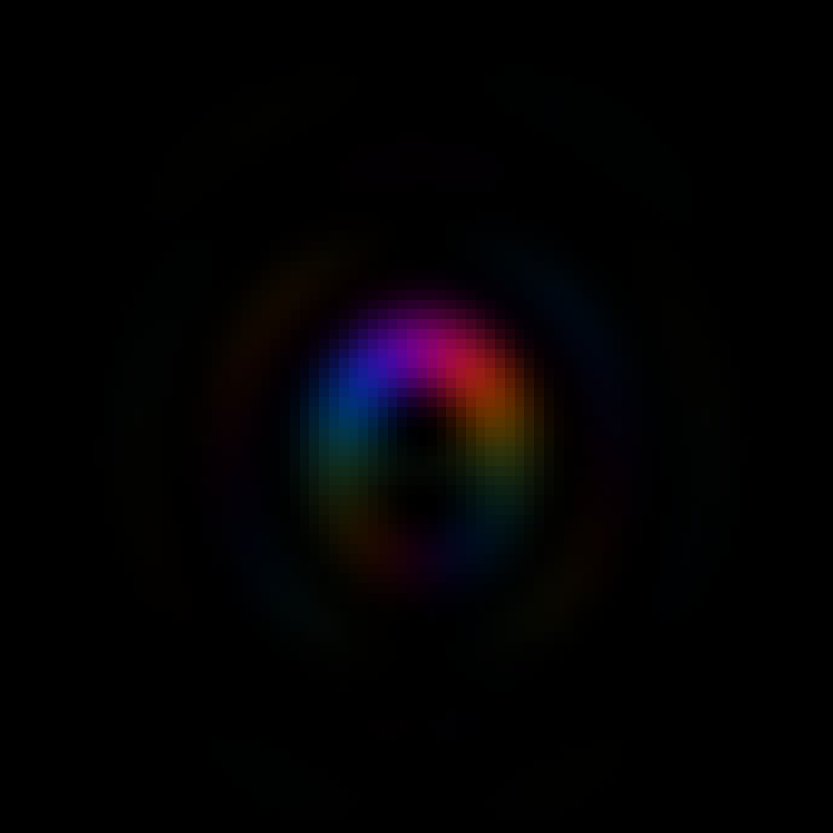}  \\

Input $\ell=3$ 
\includegraphics[width=1.5cm]{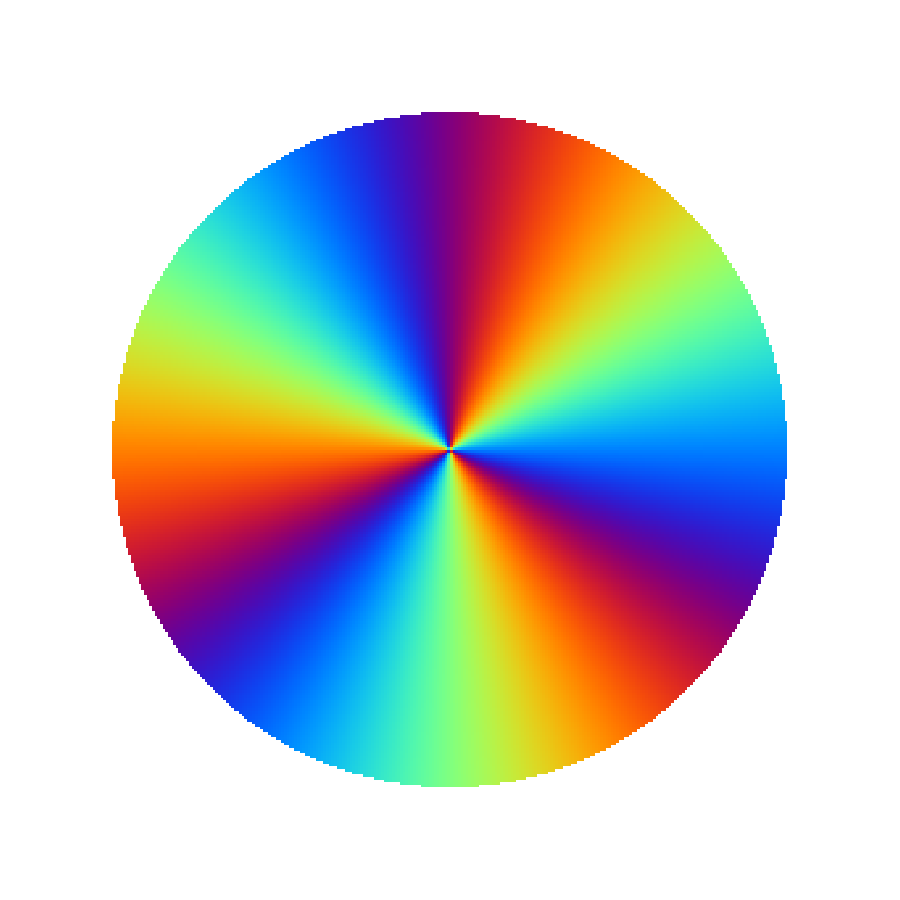}&  \includegraphics[width=3cm]{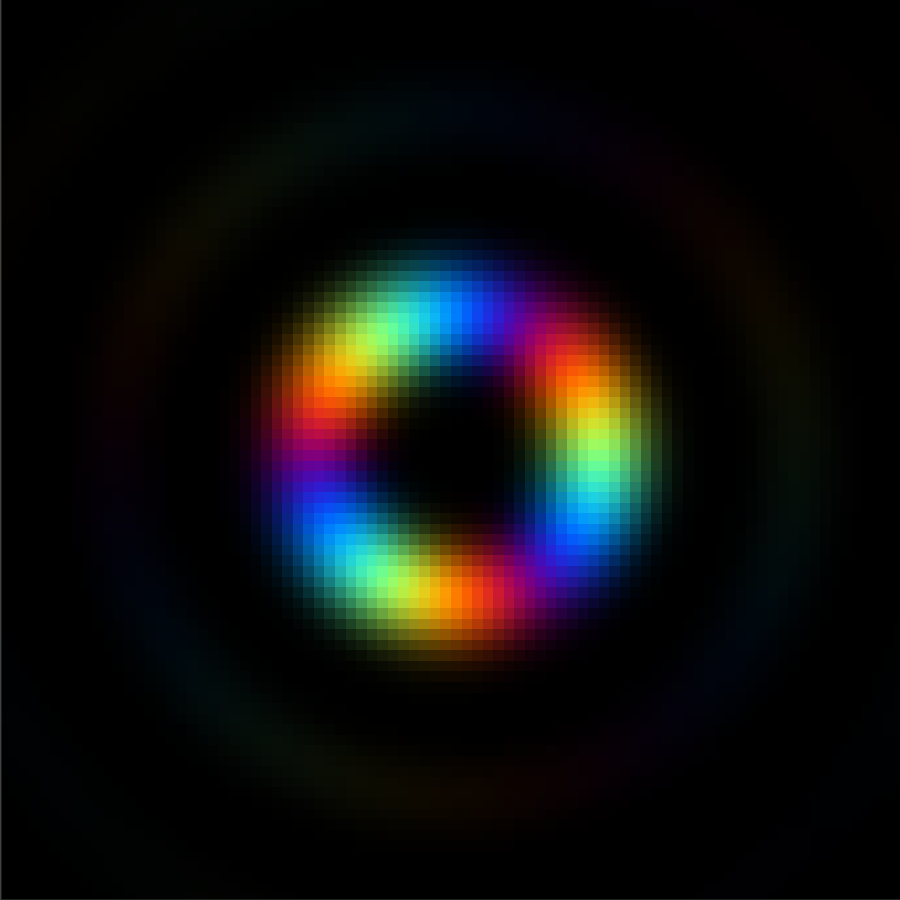}   &   \includegraphics[width=3cm]{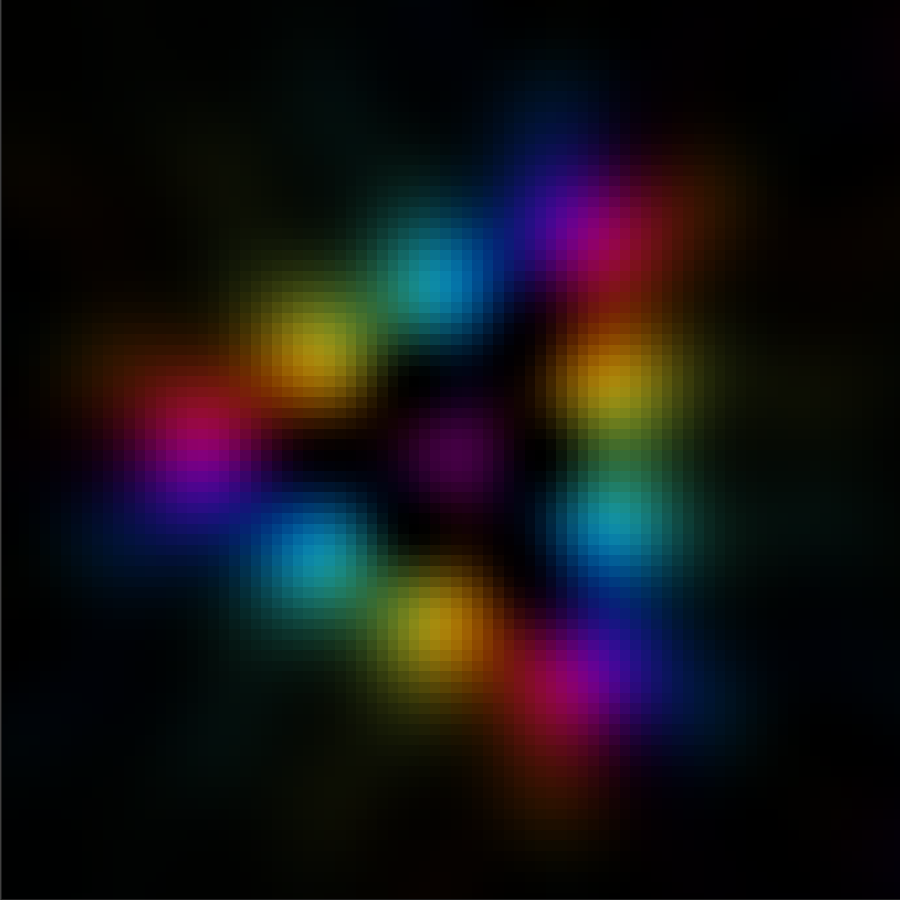}  &   \includegraphics[width=3cm]{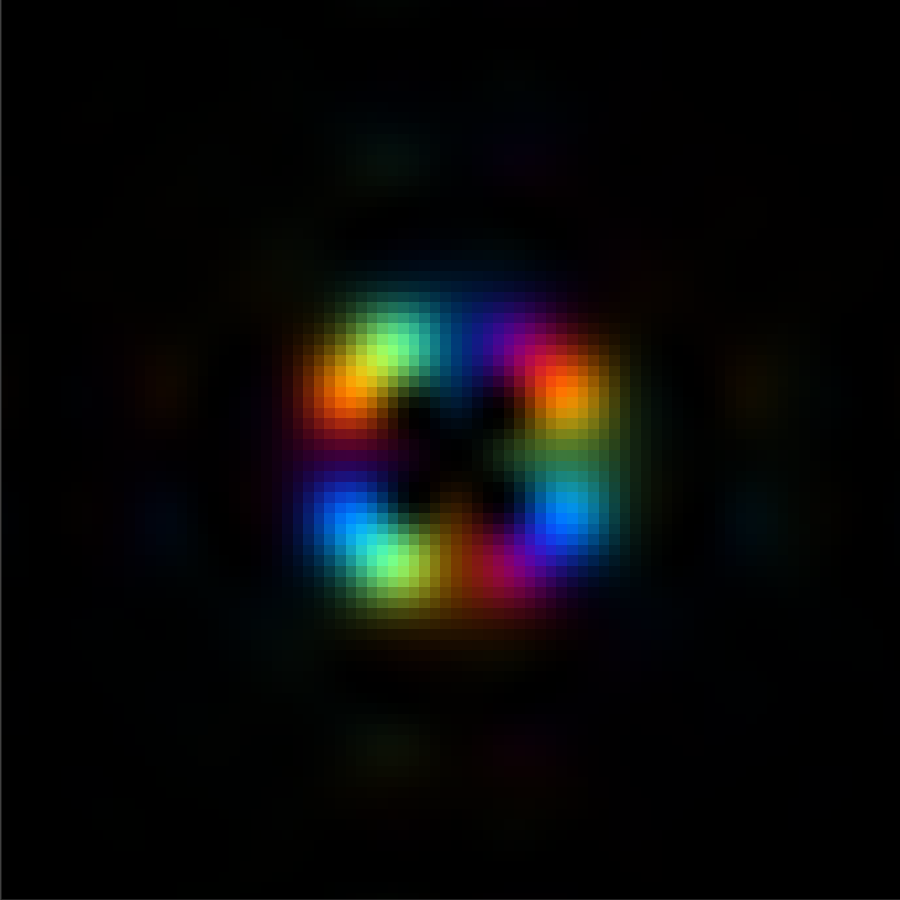}  &  \includegraphics[width=3cm]{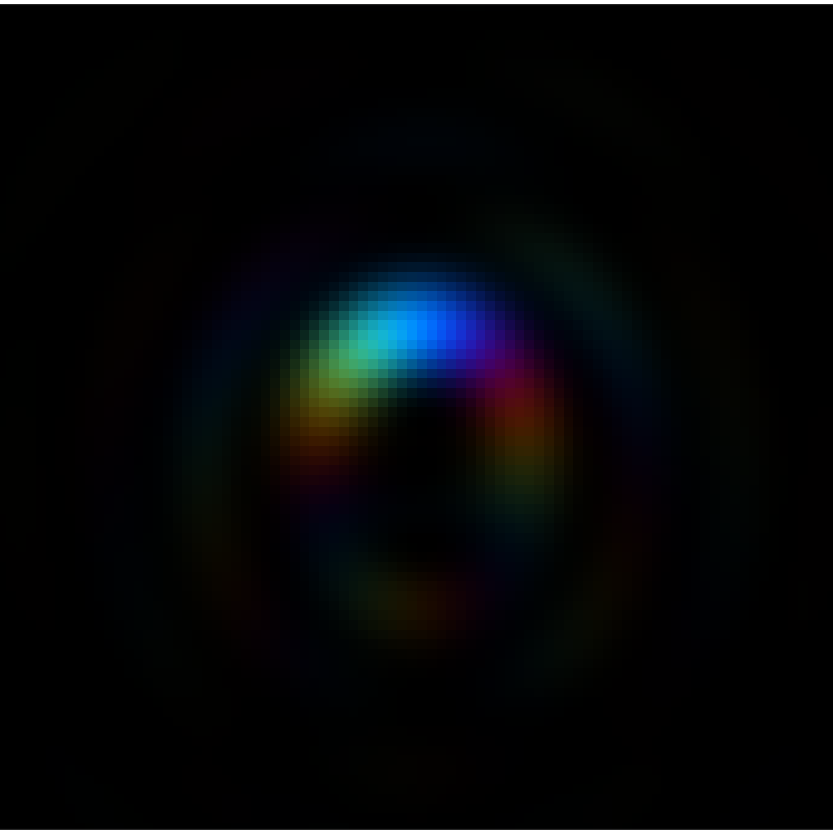}  \\
\end{tabular}
\caption{\label{TheoFFtable} Simulated apertures and far-field intensity patterns resulting from the vortex beams of different order. The $0$-$2 \pi$ phase colour map is illustrated in the $\ell=1$ incoming vortex probe. The $+$ markers in the apertures highlight the centre of the incoming vortex beam. The apertures have edge lengths, or diameters of $2 \, \mu$m.}
\end{table}

Firstly, the far-field results of illuminating a series of geometrical apertures were simulated. The chosen apertures were a centred circle, an equilateral triangle, a square, and a circle displaced by half a radius.
In Fig. \ref{TheoFFtable}, these apertures are illustrated, along with figures representing the intensity (brightness) and phase (hue) of the wave in the far-field of the aperture, when illuminated with the vortex beams of different order. The vortex beam was modelled as a beam resulting from the far-field diffraction of a forked holographic mask, with the size chosen such that the peak intensity ring of the vortex approximately matched the size of the aperture.

\subsection{Experimental Results \label{subs:Exp}}
 \begin{table}
 \centering
\begin{tabular}{>{\centering\arraybackslash}m{1.8cm} >{\centering\arraybackslash}m{3cm} >{\centering\arraybackslash}m{3cm} >{\centering\arraybackslash}m{3cm} >{\centering\arraybackslash}m{3cm}}
Apertures &  \includegraphics[width=3cm]{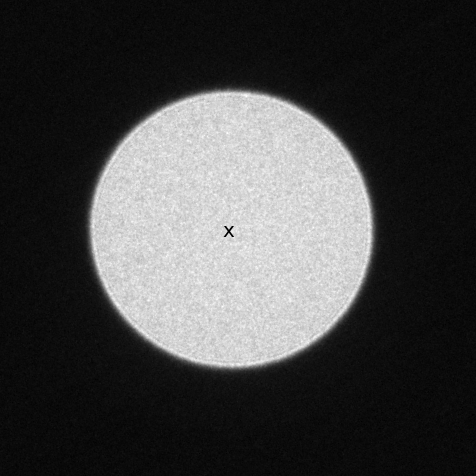}   &   \includegraphics[width=3cm]{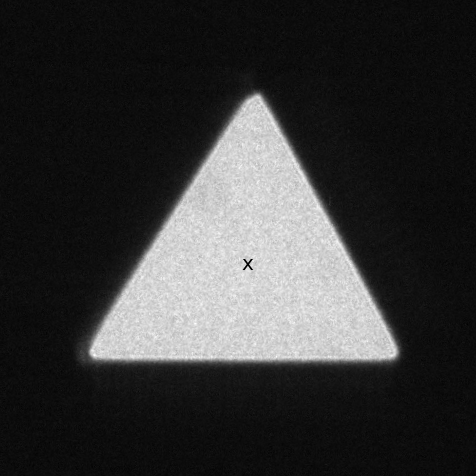}  &   \includegraphics[width=3cm]{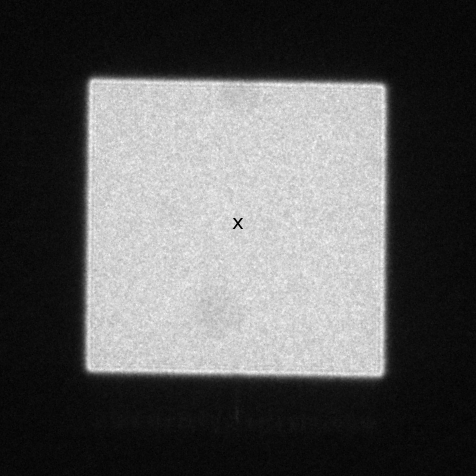} &   \includegraphics[width=3cm]{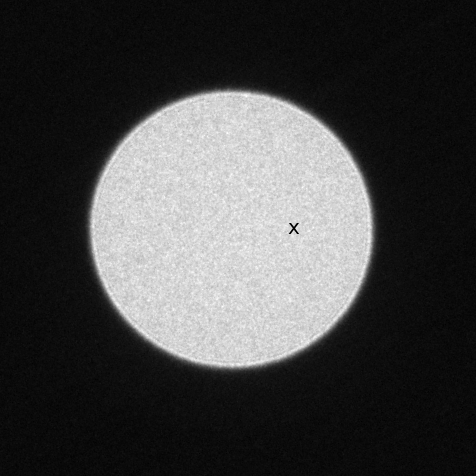}  \\

$\ell=0$ &  \includegraphics[width=3cm]{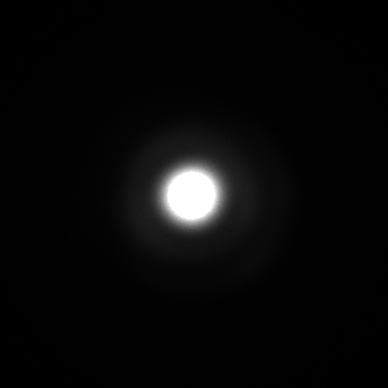}   &   \includegraphics[width=3cm]{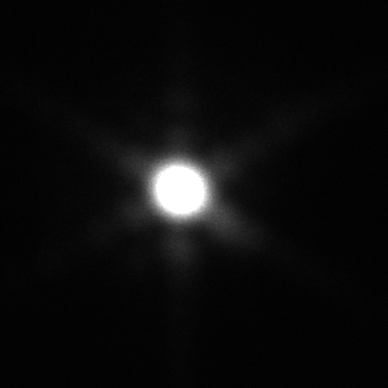}  &   \includegraphics[width=3cm]{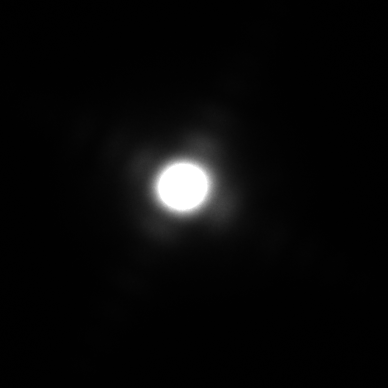}  &  \includegraphics[width=3cm]{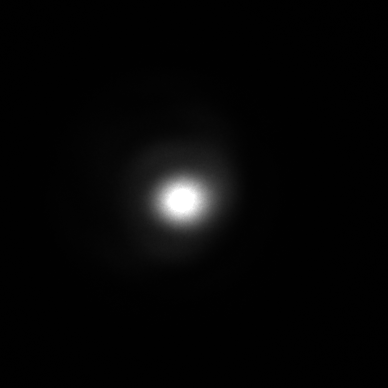}  \\

$\ell=1$ &  \includegraphics[width=3cm]{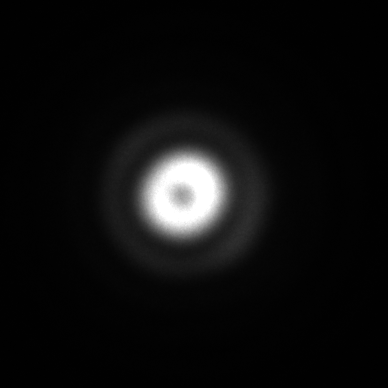}   &   \includegraphics[width=3cm]{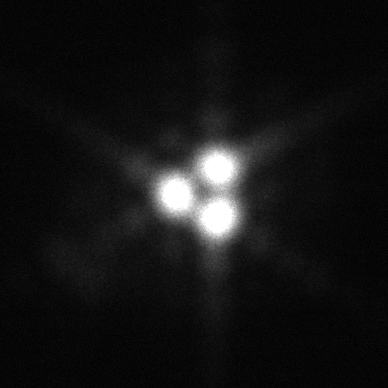}  &   \includegraphics[width=3cm]{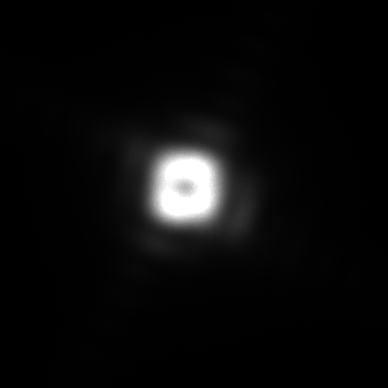}  &  \includegraphics[width=3cm]{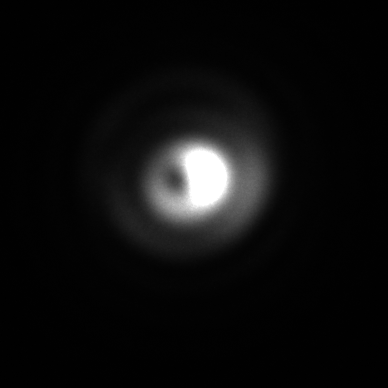}  \\

$\ell=2$ &  \includegraphics[width=3cm]{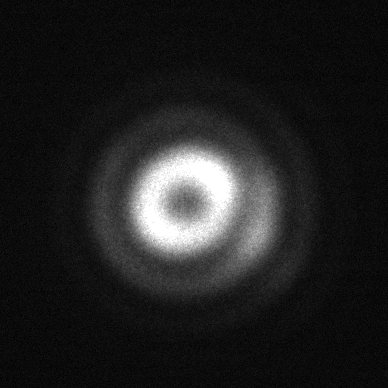}   &   \includegraphics[width=3cm]{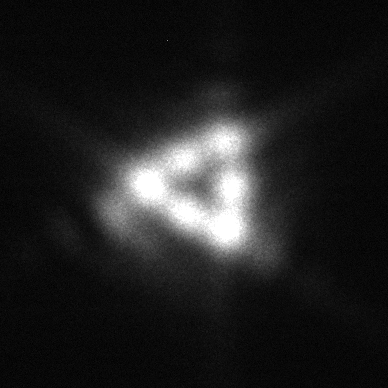}  &   \includegraphics[width=3cm]{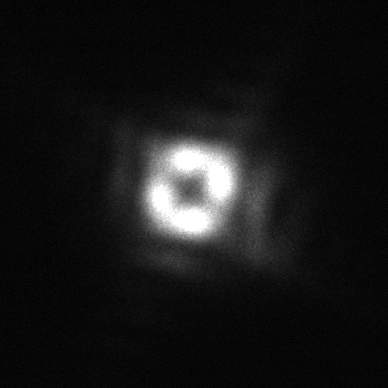}  &  \includegraphics[width=3cm]{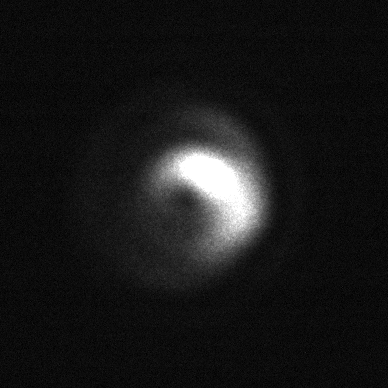}  \\

$\ell=3$ &  \includegraphics[width=3cm]{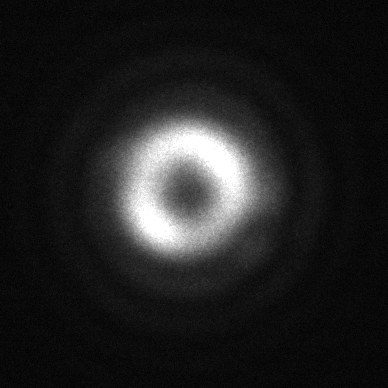}   &   \includegraphics[width=3cm]{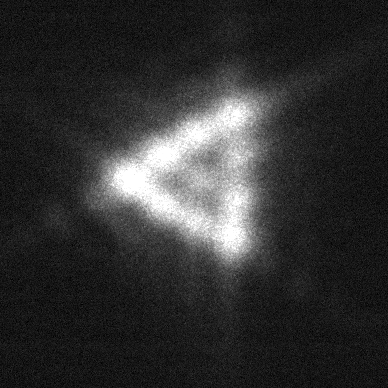}  &   \includegraphics[width=3cm]{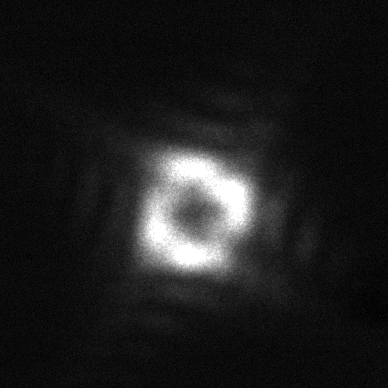}  &  \includegraphics[width=3cm]{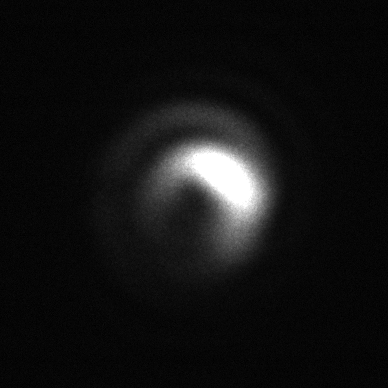}  \\

\end{tabular}
\caption{\label{ExpFFtable} Experimental apertures and far-field intensity patterns resulting from vortex beams of different order.  The $X$ markers in the apertures highlight the centre of the incoming vortex beam. The apertures have edge lengths, or diameters of $2 \, \mu$m.}
\end{table}
 The corresponding experiments were then performed in an aberration-corrected FEI Titan$^3$ microscope operating at $200$~keV, resulting in the data illustrated in Fig. \ref{ExpFFtable}.
 The forked aperture was located in the condenser plane and the geometric apertures were in the selected-area plane.
 
 Comparing the simulated and experiemental far-field images, as shown in Fig. \ref{TheoFFtable} and Fig. \ref{ExpFFtable}, we can see a very good agreement in the intensity profiles. The results begin to diverge in the $\ell=3$ data set, as these beams pass at a higher angle to the optical axis through the microscope from the forked mask and thus are more sensitive to the residual aberrations.
 
 The trivial $\ell=0$ cases simply show the intensity of a Fourier-transformed approximately-Gaussian beam, apodized by these apertures \cite{hecht1998hecht}. 
 The $\ell=0$ triangle case begins to reveal the discrete symmetry. Friedel's law points out that the Fourier transform of a real function is centrosymmetric, and thus the intensity pattern of diffraction from a triangular aperture will have six-fold symmetry.
 However, when the conditions for Friedel's law are broken, such as by a non-real, or phase-structured incoming wave (such as in the $| \ell | \geq 1 $ cases),the diffraction pattern is no longer compelled to be centrosymmetric \cite{juchtmans2015using}, and the diffraction pattern can then reveal more directional information about the probe.
 For example, \citeauthor{mourka2011visualization} established that diffraction of an optical vortex beam from an aperture without $180 ^{\circ}$ rotational symmetry can determine its handedness, through interaction of the OAM-induced rotation, with the Gouy phase \cite{mourka2011visualization, mourka2014probing}.
 It has also been shown, in both light optics and electron microscopy,  that diffraction of a single-ringed vortex beam from a triangular aperture enables a quantification of the topological order of the vortex beam, by counting the $| \ell| +1$ intensity lobes found along one edge of the resultant triangular pattern of lobes \cite{hickmann2010unveiling, guzzinati2014measuring}. 
 A closer look at the intensity profile of the $\ell = 3$ triangle case in figure \ref{ExpFFtable}, reveals a faint intensity lobe on-axis, even in the idealised simulation in figure \ref{TheoFFtable}. This is unusual in coherent vortex studies, and warrants further attention.
 A related unusual effect can be seen in the $\ell = 3$ square case, where in the centre of the intensity profile, five dark spots can be seen, arranged as on a die.
 
 To investigate the cause of these effects, the simulations were further developed to highlight the vortex cores and determine their handedness.
In a small, pixellated loop around a vortex core, phase aliasing errors can quickly creep in when dealing with higher order vortices (with their rapid phase changes). To avoid this complication, images shown in figure \ref{TheoFFvortexTable} only differentiate between the left-  or right- handedness of the vortices, and not their topological order.
 
 \begin{table}
\centering
\begin{tabular}{| >{\centering\arraybackslash}m{1.8cm} | >{\centering\arraybackslash}m{3cm} | >{\centering\arraybackslash}m{3cm} |  >{\centering\arraybackslash}m{3cm} | >{\centering\arraybackslash}m{3cm}|}
\hline
Apertures  &  \includegraphics[width=3cm]{figRoundAperture}   &   \includegraphics[width=3cm]{figTriangleAperture}  &   \includegraphics[width=3cm]{figSquareAperture} &   \includegraphics[width=3cm]{figRoundShiftedAperture}  \\ \hline

$\ell=0$ &\rule{0pt}{4ex}   \includegraphics[width=3cm]{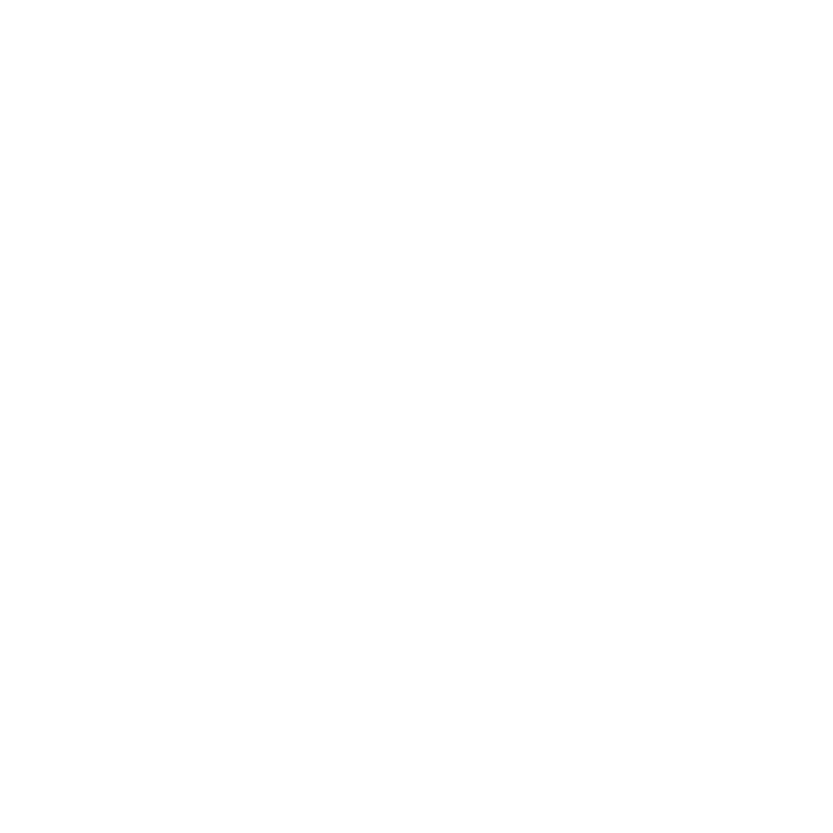}      &   \includegraphics[width=3cm]{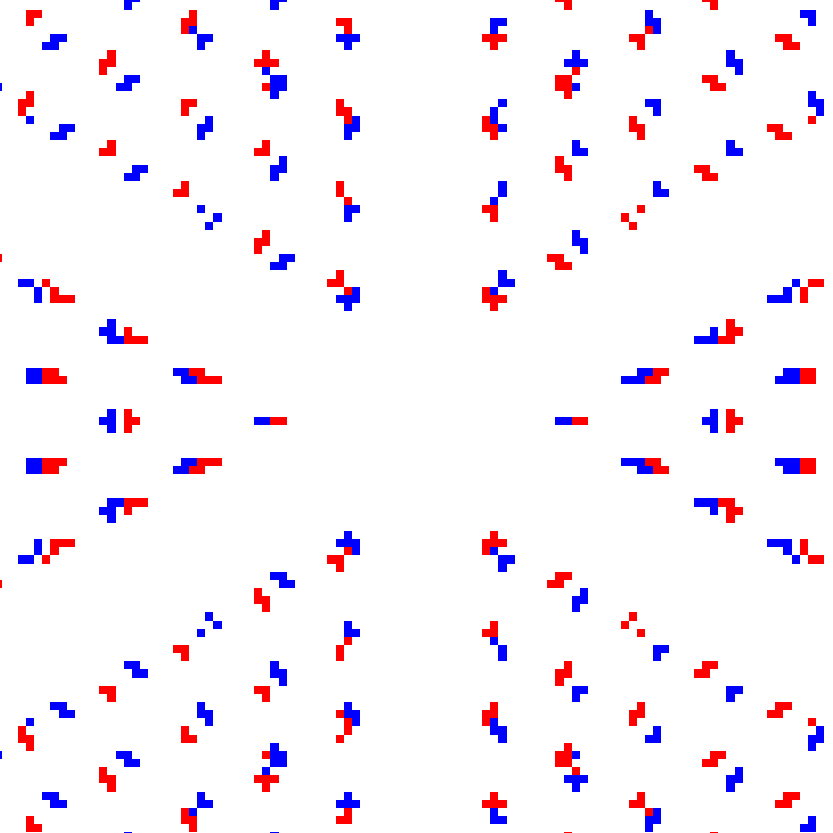}  &   \includegraphics[width=3cm]{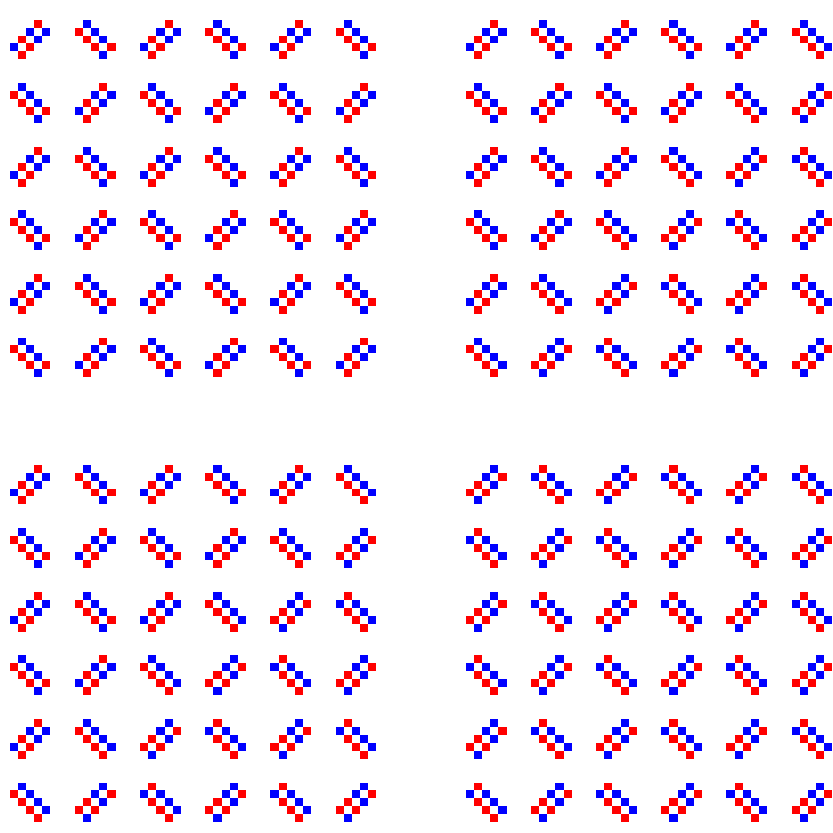}  &  \includegraphics[width=3cm]{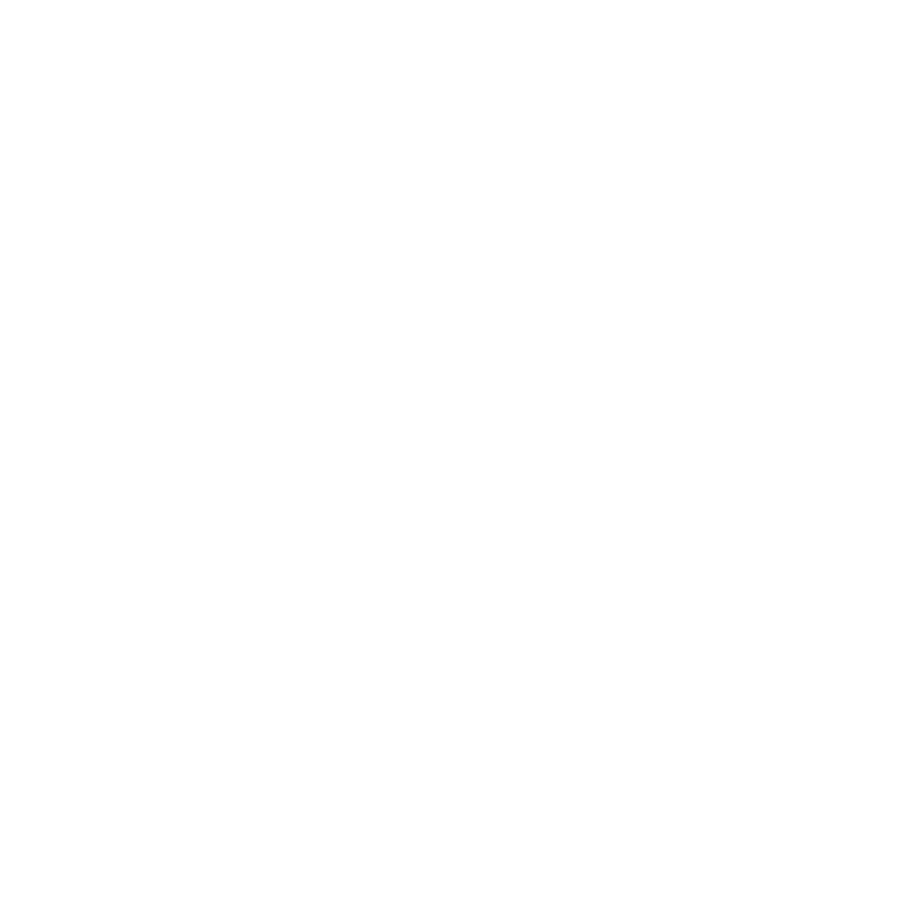}  \\ \hline

$\ell=1$ & \rule{0pt}{4ex}   \includegraphics[width=3cm]{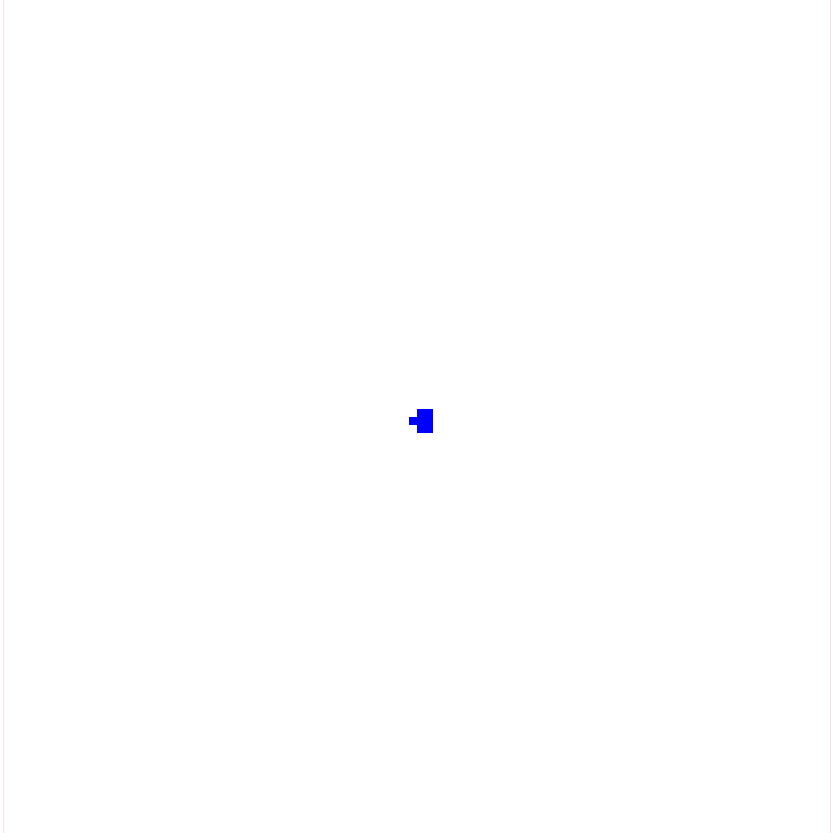}   &   \includegraphics[width=3cm]{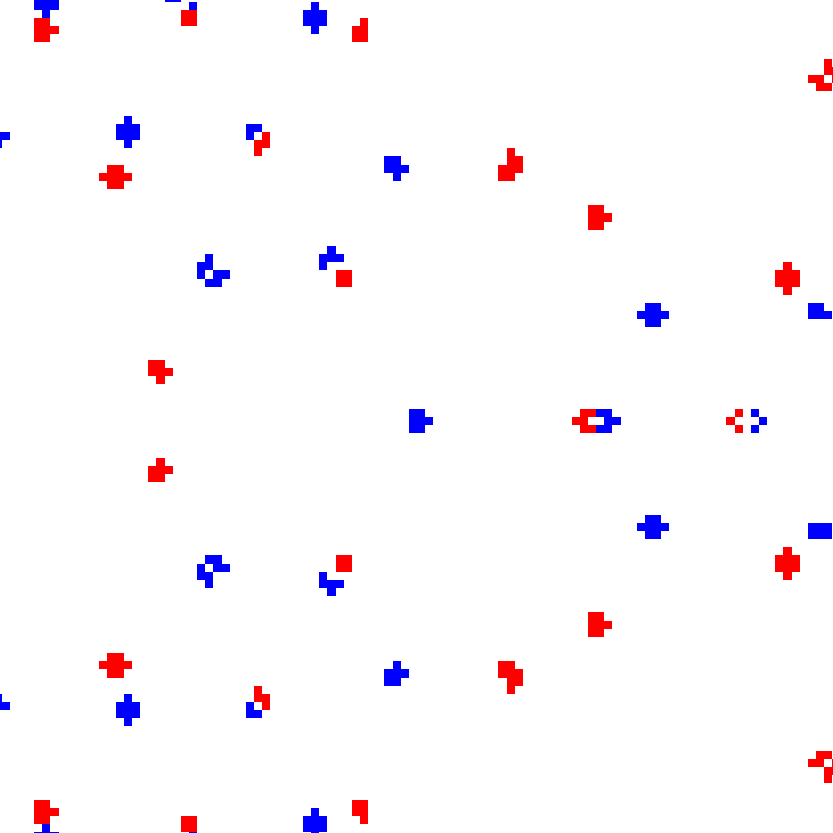}  &   \includegraphics[width=3cm]{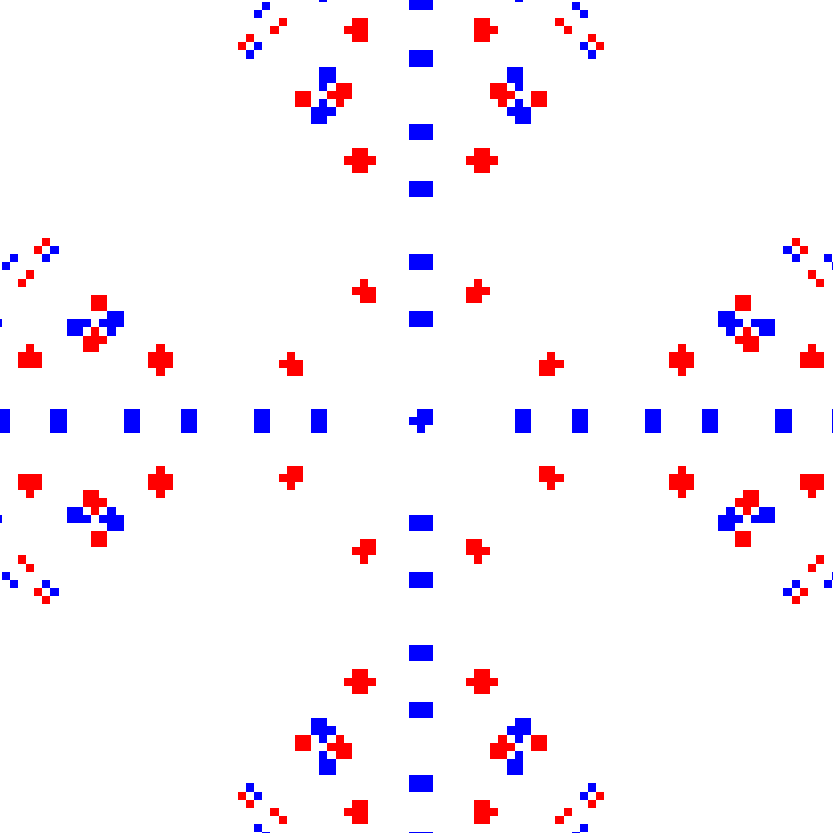}  &  \includegraphics[width=3cm]{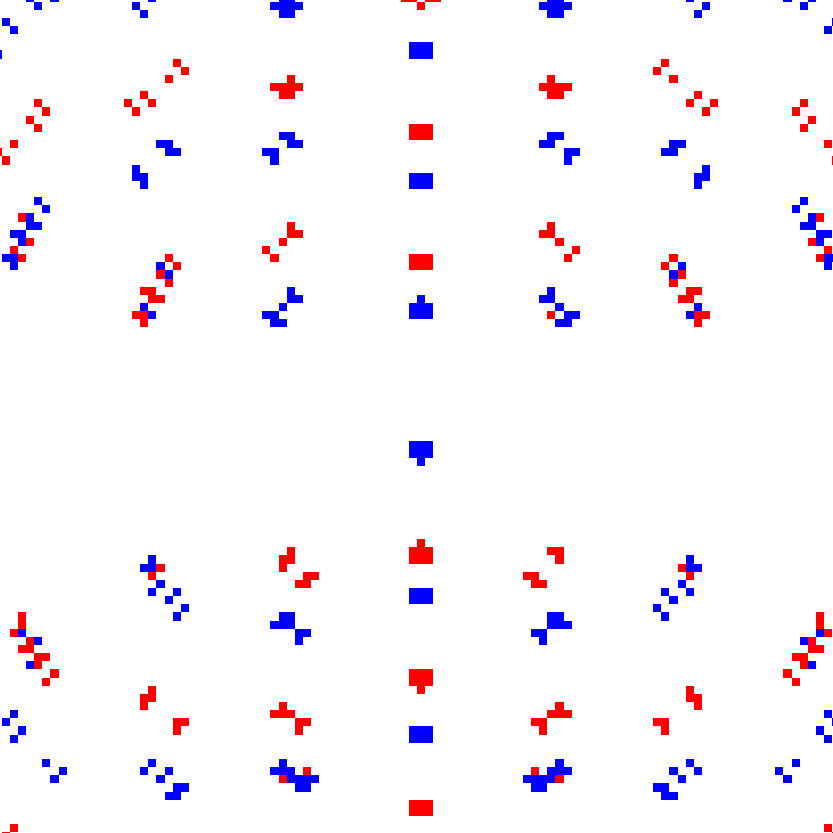}  \\ \hline

$\ell=2$ &  \rule{0pt}{4ex}  \includegraphics[width=3cm]{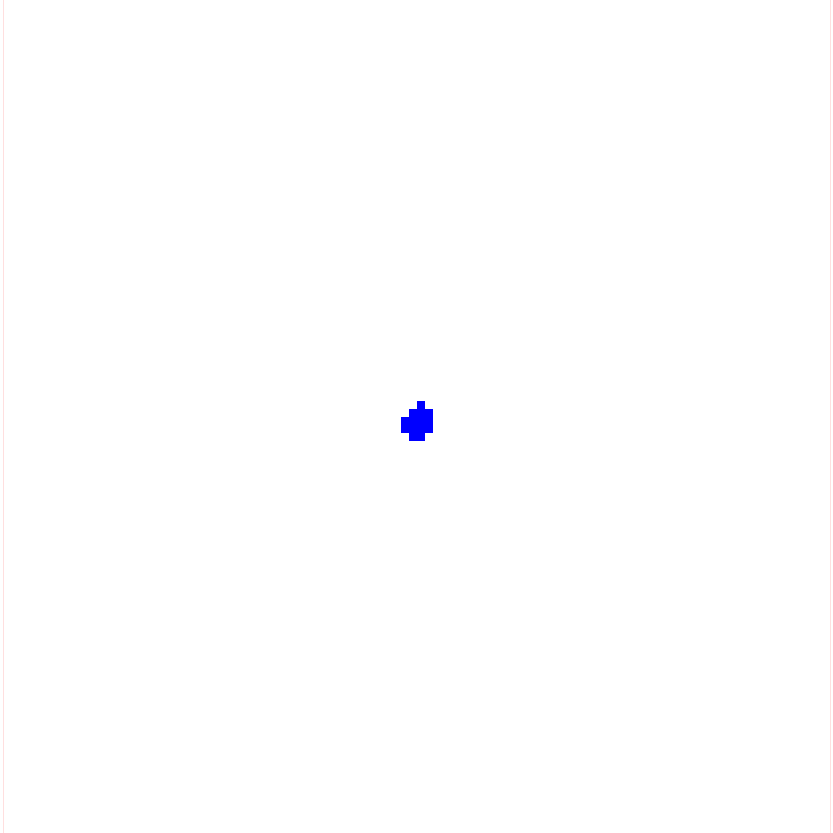}   &   \includegraphics[width=3cm]{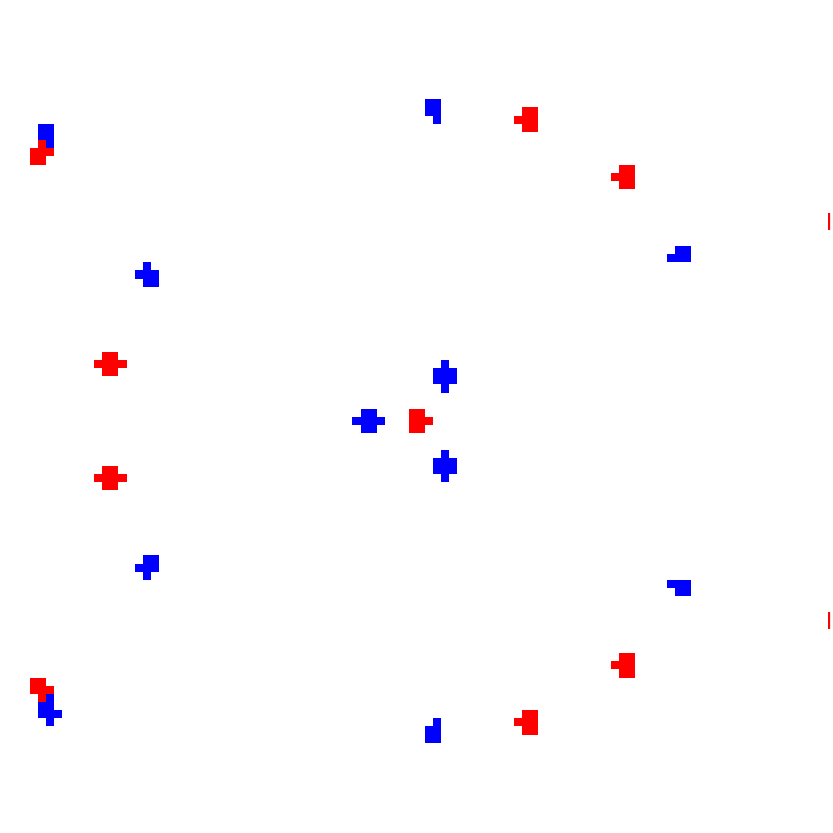}  &   \includegraphics[width=3cm]{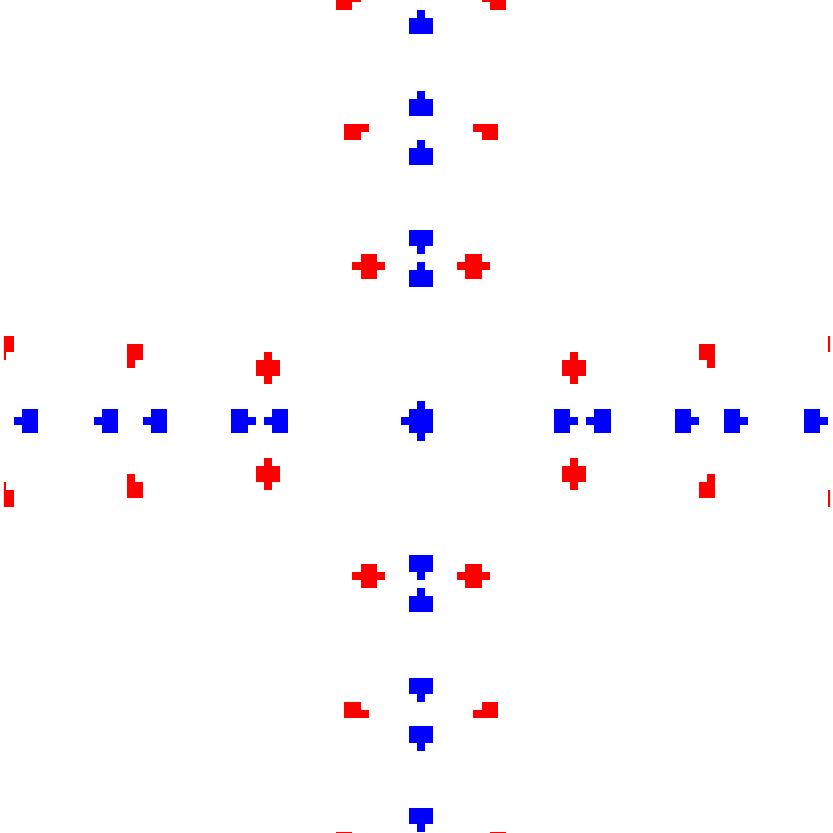}  &  \includegraphics[width=3cm]{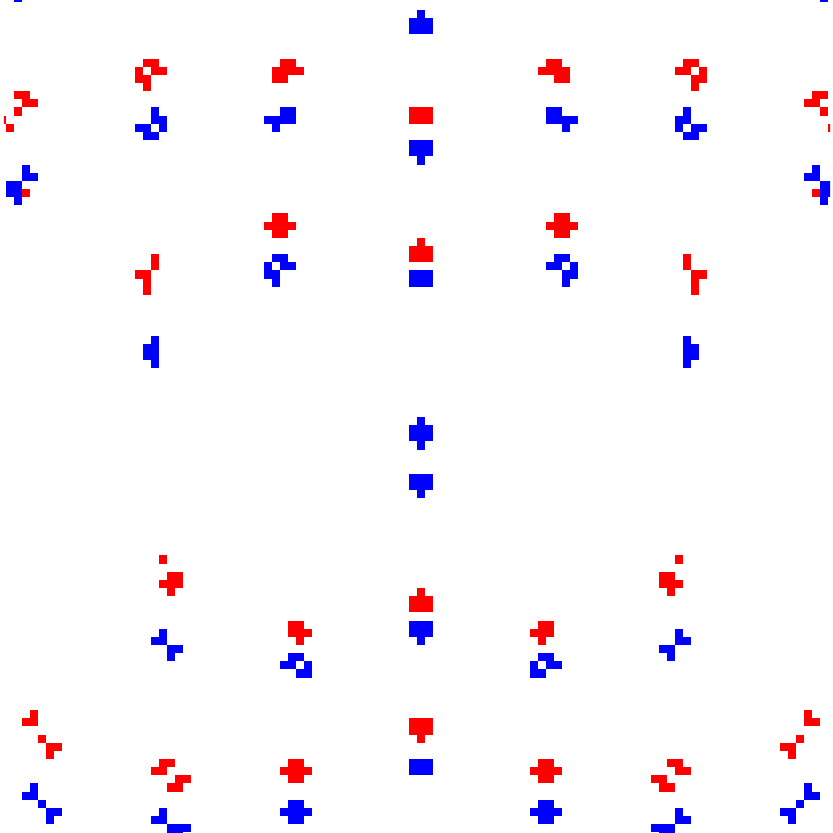}  \\ \hline

$\ell=3$ & \rule{0pt}{4ex}   \includegraphics[width=3cm]{figVORTRoundApertureFT_l3}   &   \includegraphics[width=3cm]{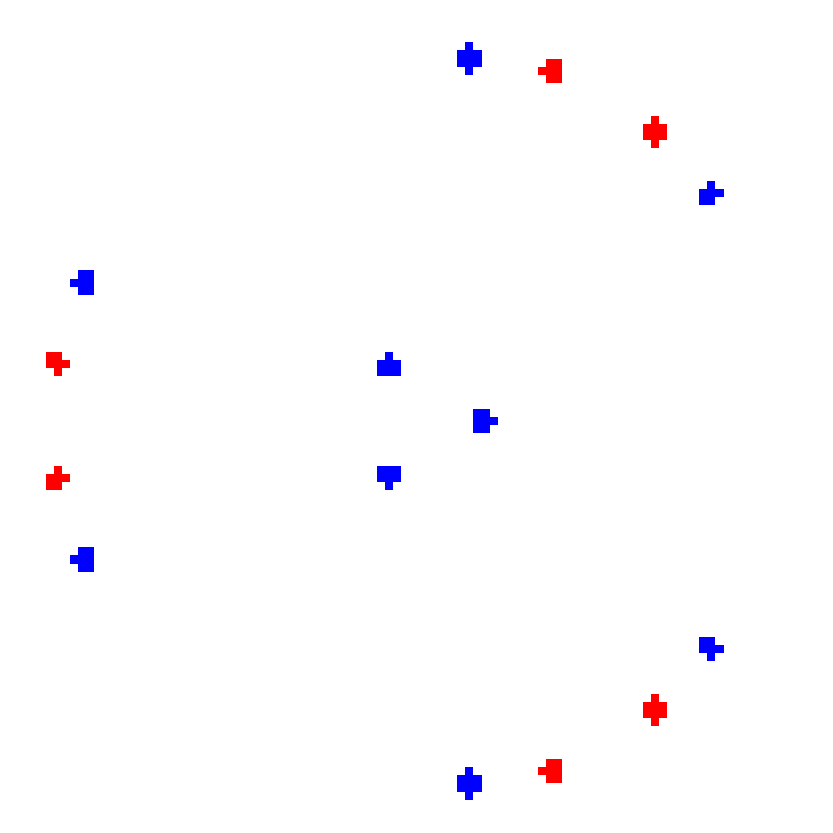}  &   \includegraphics[width=3cm]{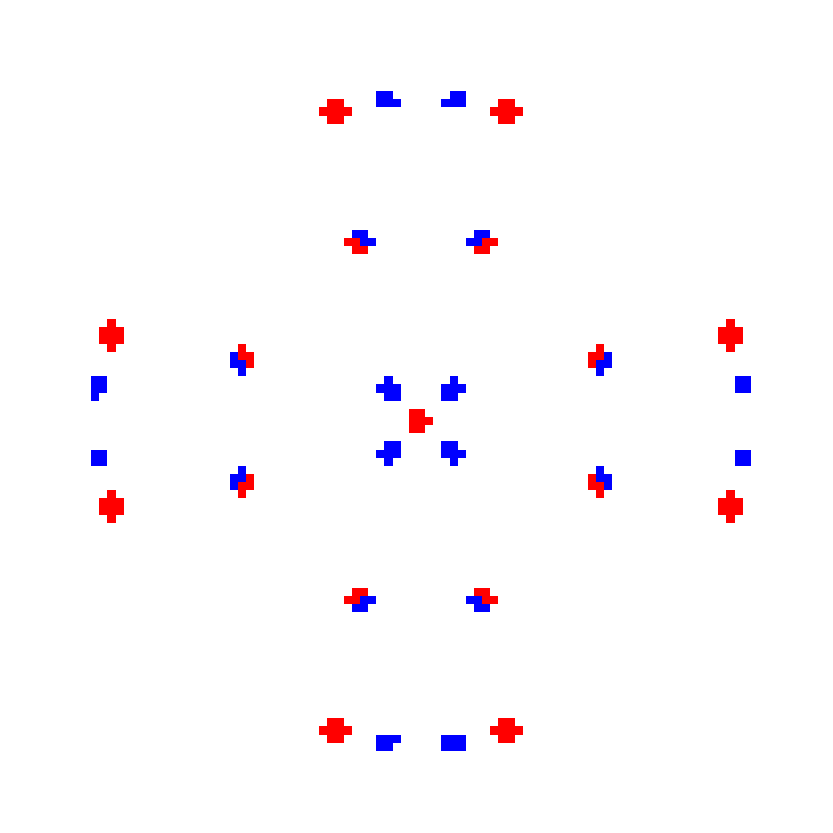}  &  \includegraphics[width=3cm]{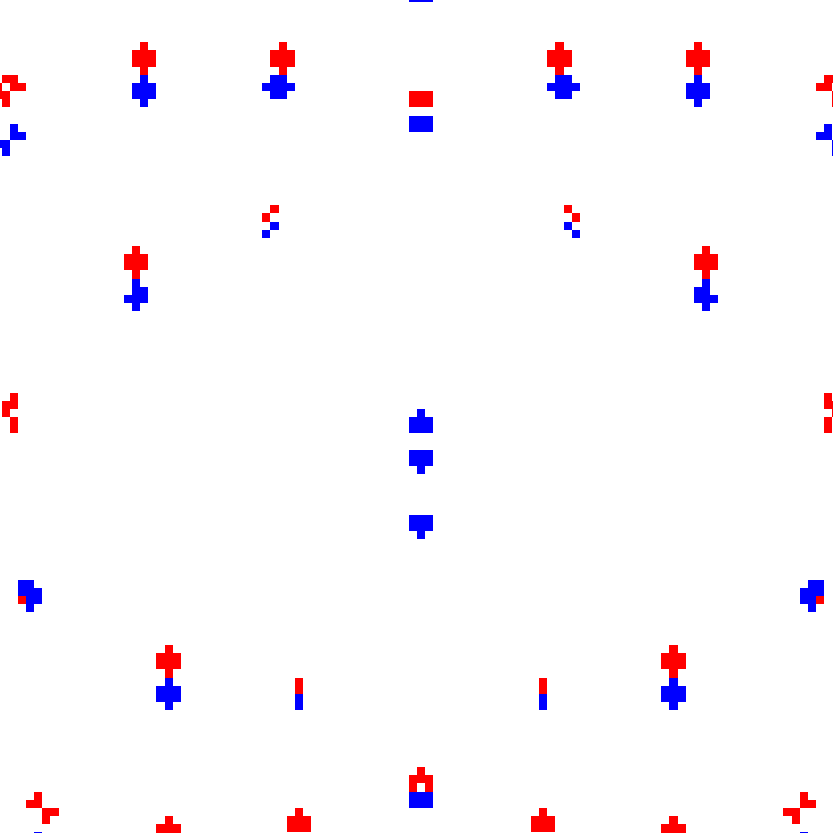}  \\ \hline

\end{tabular}
\caption{\label{TheoFFvortexTable} Simulated apertures and far-field vortex patterns resulting from the vortex beams of different order.  The $+$ markers in the apertures highlight the centre of the incoming vortex beam. Blue corresponds to a positive (right-handed) vortex core, while red corresponds to a negative (left-handed) vortex core.}
\end{table}

In the first column of figure \ref{TheoFFvortexTable}, we see the standard behaviour of the unperturbed beam, the single vortex core of each beam remains stable and on-axis, as the cylindrical symmetry permits OAM-eigenstates \cite{andrews2012angular}. There are phase jumps of $\pi$ between each ring in the tails of the Airy disc, but these do not affect the vortex behaviour in this cylindrically-symmetric case.

However, in the triangle and square cases (second and third columns of figure \ref{TheoFFvortexTable}) vortices are abundantly present in the $\ell = 0$ far-field profiles. This is due to the interference between the different sets of Fresnel fringes propagating from the aperture sides \cite{keller1962geometrical} in such a way as to produce arrays of $(+1, -1)$ vortices, similarly to the vortex lattices produced by  \citeauthor{niermann2014creating} \cite{niermann2014creating}.
 
In addition to the many vortex-antivortex pairs created in the outer region of the pattern, in some of the higher order cases, we can also note interesting splitting behaviours of the central vortex.
The splitting only occurs when the  order of the vortex is greater than half of the order of the rotational symmetry of the aperture - this inequality was first noted by \citeauthor{ferrando2005discrete} \cite{ferrando2005discrete}. We see evidence of this in the $\ell=2$, square aperture case. It is unusual for a high order vortex not to degenerate in a non-circularly symmetric system, but as in this case, $|m|\leq \frac{n}{2}$, following the notation of \citeauthor{garcia2009angular}, where integer $m$ is the angular momentum of the input beam, and $n$ is the order of rotational symmetry  of the system \cite{garcia2009angular}.
In the case where the vortex order matches the geometrical, rotational symmetry of the aperture, such as in the $\ell=3$ beam in a triangle, we see the vortex splits into $\ell$ distinct vortices of order one. 
In cases where the vortex order does not match the aperture symmetry, but does surpass \citeauthor{ferrando2005discrete}'s inequality, such that the vortex core must degenerate, additional $(+1, -1)$ vortex pairs can be created such that both topological and geometrical requirements are fulfilled. This effect we see clearly in the $\ell=2$ triangle case, and the $\ell=3$ square case where in the central region of the diffraction pattern, we can see extra vortex cores in figure \ref{TheoFFvortexTable}.
The off-centred circle however, shows a rather different vortex  structure, due to the complete absence of rotational symmetry in the system. In this case, the higher-order vortices split without creation of additional vortices in the central region of the diffraction pattern. There are many  $(+1, -1)$ vortex pairs in the surrounding area, and particularly along a line perpendicular to the aperture-shift direction (half of the Gouy rotation for a full image-to-image propagation \cite{guzzinati2013observation}).
The rotations seen in the experimental case do not match this, due to the experimental difficulties in positioning the aperture.
We note the interesting parallel between this infinite line of vortices, and the vortices resulting from a fractional phase step \cite{berry2004optical}.
  
 \subsection{Propagation series}
 
 The far-field structures certainly have interesting and unusual structures, which are structurally very different from the input wave-fronts. For this reason, we decided to also investigate the wave-front structure upon propagation.

A simulation was built around a thin-lens model, propagating from the image plane (showing the aperture outline) towards the back-focal plane (containing the far-field image).
This range is too large to be numerically modelled with a simple Fresnel propagator, without encountering the sampling issues common in such propagation methods \cite{healy2009sampling}. A modified linear canonical transform (LCT) algorithm was used instead, wherein the fractional Fourier transform allows us to sample the full range of the propagation volume.
The modelled system is an $\ell=3$ Laguerre-Gaussian vortex beam, illuminating a square aperture. This example was selected as it displays most of the interesting features in one system.

\begin{table}
\centering
\begin{tabular}{ccc}
\begin{overpic} [width=4cm]{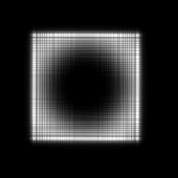}  \put(-15,25){\rotatebox{90}{{\large Simulation}}} \end{overpic}  \vspace{4 mm}  \hspace{4 mm}        
&  \includegraphics[width=4cm]{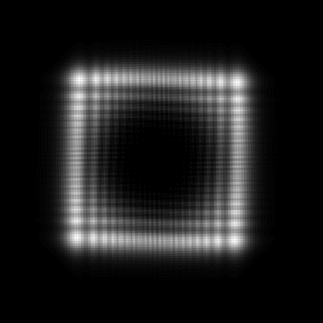} \hspace{4 mm} & \includegraphics[width=4cm]{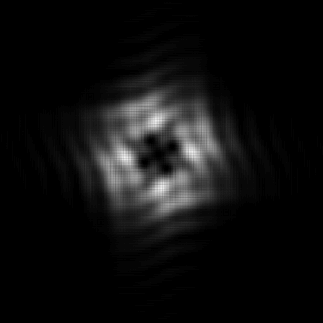}\\
\begin{overpic} [width=4cm]{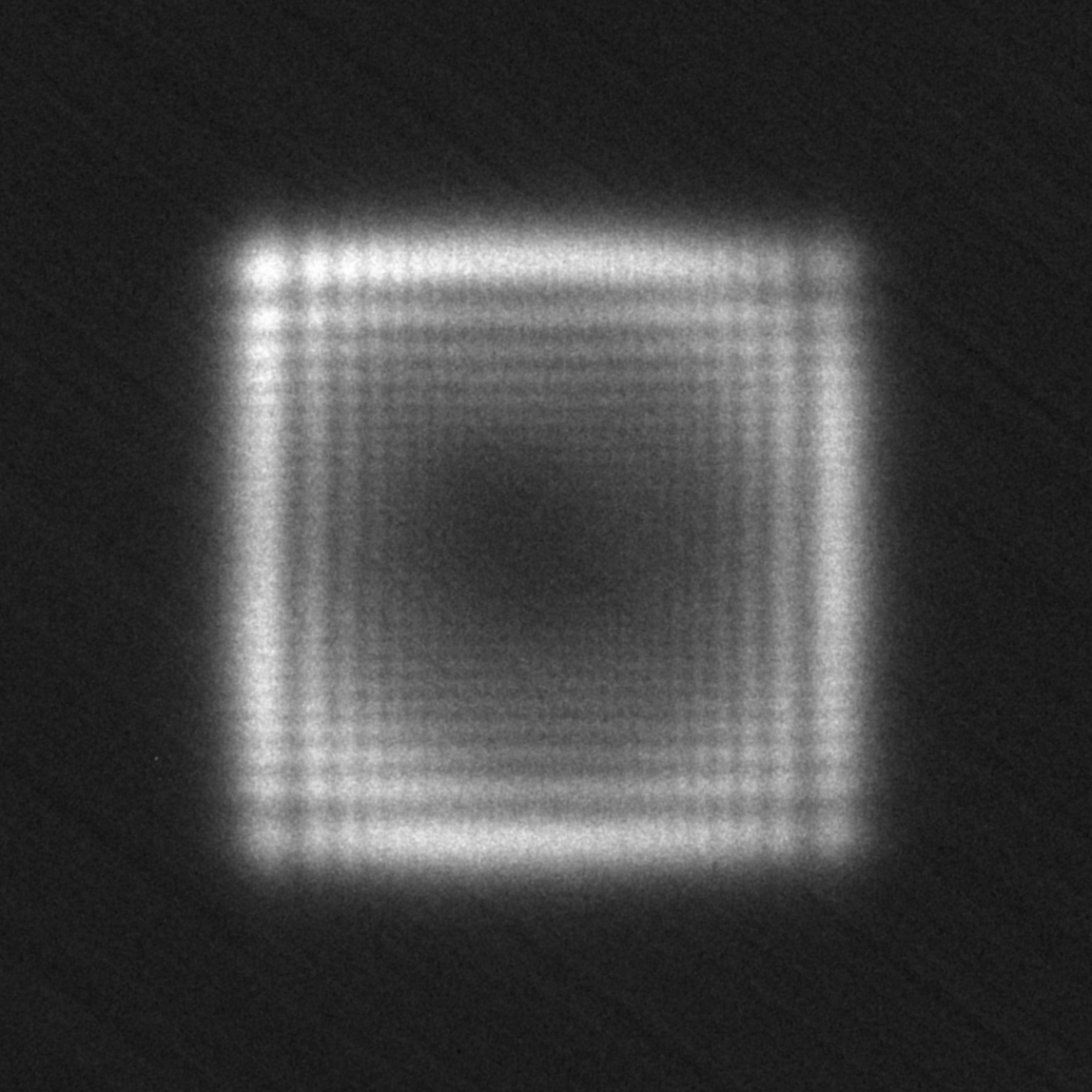}  \put(-15,23){\rotatebox{90}{{\large Experiment}}} \end{overpic}  \vspace{4 mm}  \hspace{4 mm} &  \includegraphics[width=4cm]{ExpHalfway} \hspace{4 mm}  & \includegraphics[width=4cm]{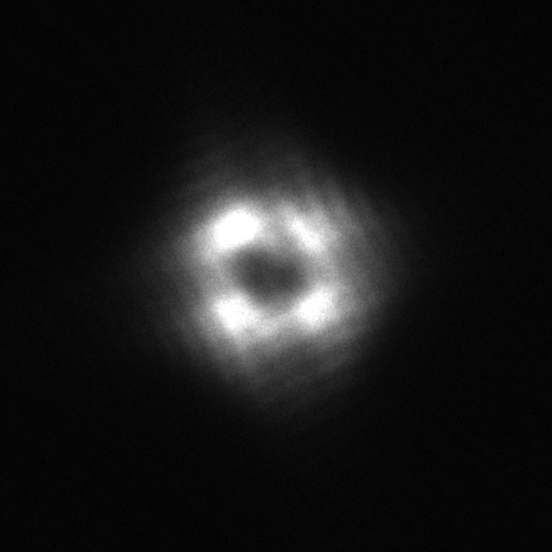}\\
\end{tabular}
\caption{\label{DFseriesTable} Propagation series of an $\ell = 3$ vortex beam after a square aperture, from near-field (left), towards the far-field (right).The upper row shows simulated data (perfect coherence); the lower row shows experimental data.}
\end{table}

Experimentally, this was performed at $300$~keV, with the same microscope set-up as described in section \ref{subs:Exp}.
The focal series is obtained by defocussing from the far-field intensity pattern, towards an image of the aperture.
This defocussing however, changes the currents in the diffraction lens, such that the principal planes, and relative position become undefined.
Subsequently, the magnification of the system changes non-linearly with the defocus. For this reason, the experimental focal series data are unscaled, and can only be compared qualitatively with the simulations.
Accordingly, the simulations are unscaled in the $(x,y)$ plane.

The simulation and experiment match closely, as can be seen in sample images from each focal series, displayed in Fig. \ref{DFseriesTable}. From the simulated series, we are also able to locate the vortex cores (using the same algorithm as for Fig. \ref{TheoFFvortexTable}), in each $(x,y)$ plane. These can then be collated over the series in $z$ by tracking isosurfaces enclosing volumes of high vorticity. This is shown in figure \ref{fig:propagation}.
   
In this figure, we can see several different regimes, as highlighted by the annotations. The incoming single vortex at the bottom of the figure is initially stable. The vortex core then splits into a complex lattice of vortices - the requirement of at least three-plane wave interference for a vortex lattice are met, with edge waves meeting the paraxial beam \cite{masajada2001optical}, leading to the initial vortex bifurcations. 
Looking further from the aperture in $z$, the created vortices annihilate in $(+1,-1)$ pairs. This is possible due to the presence of more than three plane waves \cite{oholleran2006topology}. The system cannot be expressed in only a few plane wave terms, and hence the output is very complex. The periodicity of the vortex creation-annihilation loops decreases, with the frequency of the Fresnel fringes under propagation, until the far-field is reached, transitioning approximately as the Fresnel number approaches $F=1$.

As the system reaches the far-field, the rules defined by \citeauthor{ferrando2013theory} express themselves more clearly \cite{ferrando2013theory}, with the $\langle \ell \rangle =3$ form of the beam in a square symmetry, composed of four $\ell=+1$, and one $\ell=-1$ vortex cores in the central region. In this diffraction pattern, around $86 \%$ of the wavefront is found to still be in an $\ell=3$ state (measured over the whole wave, on-axis, using the decomposition method employed in Ref. \cite{clark2013exploiting}).

Unlike the model system of \citeauthor{oholleran2009topology}, we do not see any interlinked, closed loops \cite{oholleran2009topology}. This is probably due to the predominant $z$-only momentum of the system. Unlike  \citeauthor{dennis2010isolated} we do not see any knots (excluding the unknot) \cite{dennis2010isolated}, due to the non-generic aperture shapes required to produce a knot.
   
\setcounter{figure}{5}    
\begin{figure}    
\begin{tikzpicture}
\node [anchor=west, text width=5cm, align=center] (note1) at (-4.5,9) {\large Stable far-field \mbox{topological} structure emerges};
\node [anchor=west, text width=5cm, align=center] (note2) at (-4.5,5.5) {\large Lower frequency  of vortex loop creation};
\node [anchor=west, text width=5cm, align=center] (note3) at (-4.5,4) {\large Higher frequency of vortex loop creation};
\node [anchor=west, text width=5cm, align=center] (note4) at (-4.5,2.65) {\large Undivided $\ell = 3$ vortex at input};
\node [anchor=west, text width=1cm, align=center] (note5) at (0.88,11) {\large z};
\begin{scope}[xshift=1.5cm]
    \node[anchor=south west,inner sep=0] (image) at (0,0) {    \includegraphics[width=0.6\linewidth]{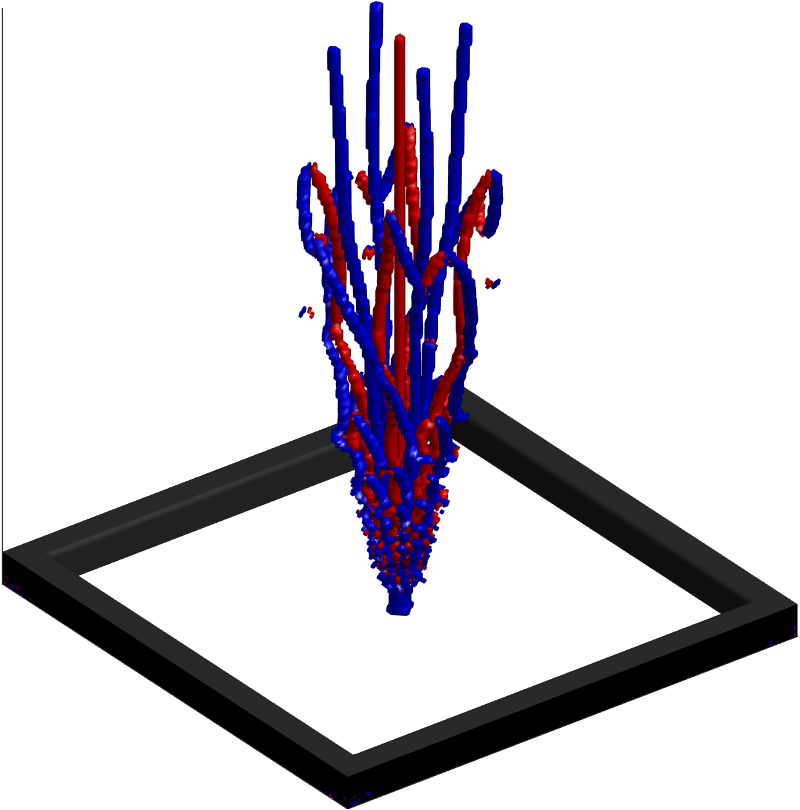}};
    \begin{scope}[x={(image.south east)},y={(image.north west)}]
        \draw [-stealth, line width=2pt, gray] (note4) -- ++(0.75,0.0);
        \draw [-stealth, line width=2pt, gray] (note3) -- ++(0.7,0.0);
        \draw [-stealth, line width=2pt, gray] (note2) -- ++(0.65,0.0);
        \draw [-stealth, line width=2pt, gray] (note1) -- ++(0.6,0.0);
    \end{scope}
\end{scope}
\end{tikzpicture}
 \caption{Tracking of the vortex cores of an $\ell=3$ vortex core, illuminating a square aperture to the far-field. Blue corresponds to a positive (right-handed) vortex core, while red corresponds to a negative (left-handed) vortex core. \label{fig:propagation}}
\end{figure}

\section{Conclusion}
We have studied the propagation of vortex cores in electron vortex beams in non-rotationally symmetric systems.
High-order vortex cores were shown to split, and it is demonstrated that a stable far-field pattern, can include more vortex-antivortex pairs than were present in the initial condition.
The propagation volume leading to such patterns was investigated for the first time, and its complexity was studied.
The vortex structure under near-field conditions was found to be exceptionally detailed.

The instability of vortices, and the ease with which they can be created under propagation, are significant contributing factors to the challenges in utilising electron vortex beams to reproducibly measure magnetic states of a material using electron energy-loss spectroscopy (EELS). The results presented in this paper suggest that to optimise vortex-EMCD, the effect of the lattice symmetry on the vortex beam needs to be minimised, perhaps through careful selection of beam convergence angle, and channelling conditions.

\section{references}
\bibliography{Clark_SymmetryConstrainedEVpropagationBIB}

\begin{acknowledgments}
L.C., A.B., G.G,  and J.V. acknowledge funding from the European Research Council under the 7th Framework Program (FP7), ERC Starting Grant No. 278510 - VORTEX.
J.V. and A.L. acknowledge financial support from the European Union under the 7th Framework Program (FP7) under a contract for an Integrated Infrastructure Initiative (Reference No. 312483 ESTEEM2).
The Qu-Ant-EM microscope was partly funded by the Hercules fund from the Flemish Government.
\end{acknowledgments}

\end{document}